\documentclass[twocolumn,amsmath,amssymb,floatfix,aps,superscriptaddress,showpacs]{revtex4} 
\pdfoutput=1
\usepackage{subfigure}
\usepackage{graphicx}

\newcommand {\vct}[1] {\mathbf {#1}}
\def\bnabla{\mbox{\boldmath $\nabla $}}

\begin{document}
\title{Confined polymer nematics: order and packing in a nematic drop}
\date{\today}

\author{Daniel Sven\v sek}
\affiliation{Department of Physics, Faculty of Mathematics and Physics, University of Ljubljana, Jadranska 19, SI-1111 Ljubljana, Slovenia}

\author{Gregor Veble}
\affiliation{School of Applied Sciences, University of Nova Gorica, Vipavska 13, P.O. Box 301, SI-5000 Nova Gorica, Slovenia}
\affiliation{Department of Physics, Faculty of Mathematics and Physics, University of Ljubljana, Jadranska 19, SI-1111 Ljubljana, Slovenia}

\author{Rudolf Podgornik}
\affiliation{Department of Physics, Faculty of Mathematics and Physics, University of Ljubljana, Jadranska 19, SI-1111 Ljubljana, Slovenia}
\affiliation{Department of Theoretical Physics, J. Stefan Institute, Jamova 39, SI-1111 Ljubljana, Slovenia}
\affiliation{Institute of biophysics, School of medicine, \\
University of Ljubljana, Lipi\v ceva 1, SI-1111 Ljubljana, Slovenia}

\begin{abstract}
We investigate the tight packing of nematic polymers inside a confining hard sphere. We model 
the polymer {\sl via} the continuum Frank elastic free energy augmented by a simple density 
dependent part as well as by taking proper care of the connectivity of the polymer chains when 
compared with simple nematics.  The free energy {\sl ansatz} is capable of describing 
an orientational ordering transition within the sample between an isotropic polymer solution 
and a polymer nematic phase. We solve the Euler-Lagrange equations numerically 
with the appropriate boundary conditions for the director and density field and investigate the orientation and density profile within a sphere. Two important parameters of the solution are the exact locations of the beginning and the end of the polymer chain. Pending on their spatial distribution and the actual size of the hard sphere enclosure we can get a plethora of various configurations of the chain exhibiting different defect geometry.
\end{abstract}


\maketitle

\section{Introduction}

Packing of DNA within simple viruses  has recently attracted a lot of attention from the physics community \cite{Gelbart1} 
since it appears that many if not all processes  connected with the bacteriophage DNA injection are governed by biologically unspecific physical mechanisms.  Cryomicroscopy of simple viruses, such as bacteriophages T7 \cite{T7}, epsilon15 \cite{epsilon15} and $\phi29$ \cite{phi29}, indicates that at elevated densities DNA appears to be wrapped into a coaxial inverse spool, with pronounced ordering and high density close to the capsid wall that both appear to decay close to the center of the capsid. Addition of polyvalent counterions such as the tetravalent spermine can induce 
a toroid formation inside the capsid reminiscent of the toroids observed {\sl in vitro} \cite{Amelie}. These toroid-like packings seem to be observed in several bacteriophages but recent studies of Leforestier et al. \cite{Livolant,domains} on cryomicroscopy of T5 bacteriophage indicate that more complicated packing geometries can also be realized such as nematic monodomains separated by defect walls that in general do not conform to the inverse spool paradigm. Within this paradigm tight packing allows DNA to act like a coiled osmotic spring piled up against the inner surface of the capsid ready to release its osmotic and elastic energy through the portal complex on docking onto a bacterial wall \cite{bustamante}.

The energetics of packaging and ejection of DNA in bacteriophages has been treated on various levels of approximation \cite{odijk1,odijk2,kindt,purohit1,purohit2,tzlil,klug1,klug2} based on the Odijk - Gelbart {\sl inverse spool
model} in which the DNA chemical potential within the viral capsid is decomposed into an interaction
part and a curvature part. This decomposition can describe the ejection of the genome reasonably when compared with osmotic stress experiments \cite{evil1,evil2}. Leaving aside simulations of the genome packing within the capsid \cite{arsuaga,spakowitz} that rely on other sets of assumptions which we will not address in what follows, the different theoretical approaches rely on the additivity {\sl ansatz} for the total free energy of DNA packing confined to a spherical capsid. They assume that the free energy is composed of two parts:  the curvature energy of DNA that is forced into the confines of the capsid, as well as  DNA-DNA interaction free energy consistent with osmotic stress experiments in the bulk \cite{osm1,osm2}. In physical terms this additivity {\sl ansatz} constitutes the basis of the inverse spool paradigm.

The form of the elastic curvature energy is known, though some recent work might indicate additional subtleties that are usually not considered \cite{nelson}. Its form, proportional to the square of local DNA curvature, follows from the standard Euler-Kirchhoff model of an elastic filament or in its continuum form from the Frank elastic energy. Though this model contains some subtle features due to the strong interhelical forces between the segments of the molecule \cite{osm1}, it nevertheless appears to be a consistent description of DNA on  mesoscopic scales \cite{rouzina}. The parameters of the Euler-Kirchhoffian model of DNA, such as its persistence length, are well established and have been measured by a variety of methods with satisfactory consensus among the results \cite{hagerman}. One can formulate the curvature elastic energy either on a single molecule level based on the Euler-Kirchhoff expression for the bending free energy or indeed on a continuous level with a volume distribution of polymer segments where the bending free energy stems from a general {\sl ansatz} of the Frank elastic free energy \cite{siber}.

The DNA-DNA interaction energy is less well known and its interpretation less well established. It can be measured directly in osmotic stress experiments \cite{methods} and can be deconvoluted into a longer ranged electrostatic contribution \cite{osm1} and a shorter ranged hydration component \cite{leikin}. Both of them have been quantified in terms of magnitudes and decay lengths \cite{podgornik3,rau}. The variable directly deduced from experiments is thus the {\sl osmotic pressure} in DNA arrays in the bulk and one can simply formulate the equation of thermodynamic equilibrium of encapsidated DNA \cite{siber} directly in terms of the measured osmotic pressure \cite{lee,podgornik-exp,rau} rather than in terms of theoretical polyelectrolyte models \cite{odijk1,odijk2,slok} or in terms of semi-empirical chemical potential expressions \cite{purohit1,purohit2,klug1,klug2}. This approach yields a consistent DNA density profile within the capsid as well as the DNA encapsidation equation of state, {\sl i.e.} the dependence of the fraction of encapsidated DNA on external osmotic pressure, that can be directly compared with experiments \cite{evil1,evil2}.

There are two major drawbacks and inconsistencies within this type of approaches. First of all these models lead to a DNA-depleted region on the central axis of the capsid but do not take into account the possibility that within this depleted region the DNA is disordered and can not be described simply and solely in terms of its elastic free energy. On top of that one usually does not solve directly for the polymer director profile within the capsid enclosure but assumes an inverse spool {\sl ansatz} form from the start. For a completely consistent description one needs a free energy {\sl ansatz} that would allow for a nematic-isotropic transition of the DNA solution as well as describe the director field configuration consistent with the distribution of the free ends of DNA within the capsid.

In what follows we will thus venture to set up a model of polymer nematic ordering within confined enclosures that would take into account the coupling between the density field and the director field of a polymer nematic as well as the possible role of the ends of the polymer chain that act as nucleators of defects in the director field. 

We start our analysis with a nematic ordering free energy appropriate for the case of long semiflexible polymers and derive the free energy contributions that are due to the well known coupling between the polymer density and director fields. This allows us to calculate the equilibrium director and density profiles within a spherical enclosure without any additional assumptions. We then investigate various packing configurations of confined polymer nematogens in order to asses the value of the inverse spool paradigm. 

Though our calculations are motivated by the problem of DNA packing within viral capsids, our results are just as relevant for a completely general case of confined semiflexible polymer ordering within spherical enclosures. While identification of our approach with the physics of capsid- confined DNA is only approximate since the persistence length of DNA is on the order of the capsid size, the appropriateness of the model would become more accurate for larger radii of the enclosure. It also addresses a previously seldom studied and thus poorly understood fundamental problem of confined polymer nematics. In any case, we are convinced that our mesoscopic approach nicely complements molecular simulations and molecular mechanics that has been used in the context of DNA packing within viral capsids before.

\section{Theoretical description}

Since DNA is a long polyelectroyte molecule its connectivity introduces additional features into a consistent continuum description of its nematic ordering that are not addressed by the standard approach of the liquid crystal physics \cite{Chaikin}. Based on previous work by Kamien et al. \cite{Kamien} we thus first introduce and formulate the concept of the {\sl polymer current} and then construct an appropriate nematic ordering free energy that we solve in confined geometry of a spherical capsid.

\subsection{Polymer ``current density''}

In a polymer nematic liquid crystal, splay deformation becomes progressively more expensive with increasing chain length, as it imposes costly local changes in polymer density \cite{Kamien,Kamien2}. In the continuous limit of long chains this coupling between the polymer density and orientational fields is described by an analogue of the continuity equation for the nematic director field $\bf n$ \cite{conteq},
\begin{equation}
	\nabla\cdot(\rho_s{\bf n}) = 0,
	\label{continuity_basic}
\end{equation}
where $\rho_s$ is the surface number density of chains crossing the plane perpendicular to the director field \cite{meyer}. 
Eq.\,(\ref{continuity_basic}) can be interpreted as the continuity condition for a polymer ``current density'' 
${\bf j} = \rho_s {\bf n}$.
The only difference between Eq.\,(\ref{continuity_basic})  and the usual continuity equation is that in this case $\bf j$ does not describe a rate (there is no time derivative involved in its definition), i.e., we are observing the number of chains perforating the perpendicular plane rather than the number of particles crossing it per unit time. For the same reason the time-dependent term (${\partial\rho\over\partial t}$ or similar, where $\rho$ is the volume density) is absent.

The analogy comes fully into life if one relaxes the condition $|{\bf n}|=1$ and takes into account the variable degree of nematic order. Let us stress that at this stage we will consider polar ordering rather than quadrupolar nematic ordering. It is not yet understood how to construct a replacement for the polymer current in case of the nematic order tensor. In principle, there is no polar ordering in a regular nematic. Yet in the case of the polymer nematic its definition is still useful. Here it can be defined because of the connectivity of individual long chains, in opposition to the  case of shorter nematogens. For sufficiently short times (such that the diffusion of ends is not effective) or if the ends are pinned, the ordering is polar, if we identify the beginning and the end of the chain. 

The description of nematic ordering will be introduced through a non-unit nematic order vector $\bf a$,
\begin{equation}
	{\bf a} = \langle\cos\theta\rangle\, {\bf n}. 
	\label{a_brief}
\end{equation}
It is defined in a given hydrodynamic volume $V$ by the expression
\begin{eqnarray}
	{\bf a} &=& {1\over\left\langle\sum_i \int_{l_i}\!{\rm d}l_i\right\rangle}\left\langle\sum_i\left(\int_{l_i}\!{\rm d}{\bf r}_i\right)\right\rangle = \nonumber\\
			&=& {1\over\left\langle\sum_i \int_{l_i}\!{\rm d}l_i\right\rangle}\left\langle\sum_i\left(\int_{l_i}\!{\rm d}l_i\cos\theta_i\right)\right\rangle \, {\bf n},
	\label{a_detail}
\end{eqnarray}
where the integrals go over the length of the $i$th chain within the hydrodynamic volume and the sum goes over all the polymer chains whose parts can be found within the hydrodynamic volume; $\theta_i$ is the angle between the local tangent on a chain and the average vicinal director field represented by $\bf n$, and $\langle\rangle$ is the thermodynamic average.

Let us now calculate the ``flux'' of chains through a surface perpendicular to the average vicinal director field $\bf n$, i.e., the number of chains perforating this surface per unit area. On average, a subunit $i$ of the chain perforates the surface if the distance between its center and the surface is smaller than ${\textstyle {1\over 2}}\int_{\ell_0}\!{\rm d}l_i\cos\theta_i$ on either side of the surface; $\ell_0$ is the subunit length.
Averaging over all chain configurations in the hydrodynamic volume we obtain
\begin{eqnarray}
	{\bf j} \cdot {\bf n}  = 
					\rho\, {\left\langle \sum_i\int_{l_i}\!{\rm d}l_i\cos\theta_i\right\rangle\over 
					\left\langle \sum_i\int_{l_i}\!{\rm d}l_i\right\rangle}\,
					{\left\langle\sum_i\int_{l_i}\!{\rm d}l_i\right\rangle\over \langle N\rangle} = \rho\, \ell_0\, \langle\cos\theta\rangle.
\label{defflux}
\end{eqnarray}
Here we introduced the volume number density of subunits $\rho$ and used the definitions (\ref{a_brief}) and (\ref{a_detail}), with $\langle N\rangle = \langle \sum_i 1\rangle$ the average number of subunits in the hydrodynamic volume. ${\left\langle\sum_i\int_{l_i}\!{\rm d}l_i\right\rangle}/\langle N\rangle$ is obviously the length of the subunit, $\ell_0$. Note that the length of the subunit may be chosen arbitrarily, in relation with the definition of the density, i.e., the product $\rho\, \ell_0$ is independent of this choice.

By definition this ``flux'' density of chains through the surface is given by the normal component of the ``current density'', $j_n = {\bf j} \cdot {\bf n}$.
We can thus define the polymer current density at any point within the sample as
\begin{equation}
	{\bf j} = \rho\, \ell_0\, {\bf a}.
	\label{current}
\end{equation}
In the infinite chain limit where there exist no chain beginnings or ends, the number of chains entering a closed surface must be equal to the number of chains exiting it, hence
\begin{equation}
	\nabla\cdot{\bf j} = 0.
	\label{div-zero}
\end{equation}
Moreover, if the chain beginnings and ends do exist, then the flux through a closed surface is nonzero: it is positive if there are more beginnings than ends within the closed surface, and negative in the opposite case. Thus, the net density of the beginnings and ends of the chains acts as a source term and we can write down a complete local continuity requirement,
\begin{equation}
	\nabla\cdot{\bf j} = \rho^{\pm},
	\label{continuity}
\end{equation}
where $\rho^{\pm}$ is the volume number density of the beginnings ($\rho^{\pm}>0$) and the ends ($\rho^{\pm}<0$) of the chains. Eq.\,(\ref{continuity}) presents a generalization of Eq.\,(\ref{continuity_basic}), taking into account variable ordering and source terms due to chain ends. It comes in the form of the usual continuity equation, except that, by construction, it does not contain the time-dependent term.

The microscopic definition of the polymer current density (\ref{current}) and the continuity equation (\ref{continuity}) can be obtained as follows (for details see e.g. \cite{Kleinert}). Let us first show that the vector
\begin{equation}
{\bf t}(\vct r) = \left\langle \sum_i \int_{l_i}\!{\rm d}l_i ~\dot{\vct r}(l_i)  \delta^3(\vct r - \vct r(l_i)) \right\rangle\label{vector-t}
\end{equation}
is just the polymer current density (\ref{current}). Above the sum goes over all the polymer molecules in the system, and the integrals go over the full length of each polymer; $\dot{\vct r}(l_i) = \frac{{\rm d}\vct r (l_i)}{{\rm d}l_i}$ is the local unit tangent vector of the chain.  
An insight into the nature of $\vct t$ is obtained by smoothing it out by integration over a hydrodynamic volume $V$ as
\begin{eqnarray}
{1\over V}\int_{V}\!{\rm d}^3 r~ \vct t(\vct r) &=&{1\over V}\int_{V}\!{\rm d}^3 r  \left\langle \sum_i  \int_{l_i} \!\!\!{\rm d}l_i ~\dot{\vct r}(l_i) ~\delta^{(3)}(\vct r - \vct r(l_i)) \right\rangle =  \nonumber\\
&=& {1\over V}\sum_i  \left\langle  \int_{l_i} {\rm d}l_i  ~\dot{\vct r}(l_i)  \right\rangle, \label{vol-t}
\end{eqnarray}
where the last sum and integral go over the chains and the segments within the hydrodynamic volume $V$, respectively.
Furthermore, by writing the microscopic subunit density of the polymer chains in the form 
\begin{equation}
\rho(\vct r) = {1\over\ell_0}\left\langle \sum_i  \int_{l_i} {\rm d}l_i ~ \delta^{(3)}(\vct r - \vct r(l_i)) \right\rangle,
\end{equation}
where $\ell_0$ is again the length of the subunit, 
we obtain that the smoothed hydrodynamic subunit number density is
\begin{eqnarray}
{1\over V}\int_{V}\!{\rm d}^3 r~\rho(\vct r) &=& {1\over V}{1\over \ell_0}\int_{V}\!{\rm d}^3 r~\left\langle \sum_i  \int_{l_i} {\rm d}l_i ~ \delta^{(3)}(\vct r - \vct r(l_i)) \right\rangle = \nonumber\\
&=& {1\over V}{1\over \ell_0}\sum_i \left\langle \int_{l_i} {\rm d}l_i \right\rangle.
 \label{vol-e}
\end{eqnarray}
Equations (\ref{vol-e}) and (\ref{vol-t}) then give back {\sl exactly} the definition of the polymer current density 
(\ref{current}) with the identification that the vector $\bf t$, Eq.\,(\ref{vector-t}), is exactly the microscopic version of the current density $\bf j$ of the polymer.

This identification of the polymer current density leads directly to the continuity requirement (\ref{continuity}) that it needs to satisfy. This can be seen by evaluating the divergence of the microscopic current density vector, obtaining
\begin{eqnarray}
& & \bnabla \cdot \vct j(\vct r) = \left\langle \sum_i \int_{l_i} {\rm d}l_i ~\dot{\vct r}(l_i) \cdot \bnabla \delta^3(\vct r - \vct r(l_i)) \right\rangle =
\nonumber\\
&=&  
- \left\langle \sum_i \int_{l_i} {\rm d}l_i \frac{{\rm d}}{{\rm d}l_i} \delta^3(\vct r - \vct r(l_i)) \right\rangle = \nonumber\\
&=& - \left( \left\langle \sum_i \delta^3(\vct r - \vct r_i(L)) \right\rangle - 
\left\langle \sum_i \delta^3(\vct r - \vct r_i(0))\right\rangle\right). \nonumber\\
~
\end{eqnarray}
Here we took into account that $\bnabla \longrightarrow - \bnabla_i$, noting that the argument of the Dirac delta function is $\vct r - \vct r(l_i)$. Also, $\dot{\vct r}(l_i) \cdot \bnabla_i = \frac{{\rm d}}{{\rm d}l_i}$. The above result can be written alternatively as
\begin{equation}
	\bnabla \cdot \vct j(\vct r) = \rho^+(\vct r) - \rho^-(\vct r),
	\label{continuity-sources}
\end{equation}
where 
$$\rho^+(\vct r)  = 
\left\langle \sum_i \delta^3(\vct r - \vct r_i(0)) \right\rangle$$ 
and  
$$\rho^-(\vct r) = 
\left\langle \sum_i \delta^3(\vct r - \vct r_i(L))\right\rangle$$ 
are the total densities of the beginnings and ends of the chains in the sample. 
Identifying $\rho^+ - \rho^- = \rho^\pm$, we have thus formally recovered Eq.\,(\ref{continuity}).

This constraint, Eqs.\,(\ref{continuity_basic}) and (\ref{continuity-sources}), on the director field in polymer nematics has been discovered by de Gennes and Meyer \cite{conteq} and bears some similarity with the differential form of the Gauss theorem in electrostatics. The divergence of the field is determined by the density of the sources. Here the orientational field $\vct j$ plays the role of the electrostatic field and the density of beginnings and ends of the chain play the role of positive and negative charges. 

In the limit of fixed nematic order and fixed density, only if there exists a proper mismatch of $ \rho^+(\vct r)$ and $\rho^-(\vct r)$ can there be a splay deformation of the polymer nematic. Eq.\,(\ref{continuity-sources}) presents a generalization of Eq.\,(\ref{continuity_basic}), taking into account variable ordering and source terms due to finite length of the polymer chains. 

\subsection{Free energy}

To determine the equilibrium configuration of the director and density fields we set up a free energy density following the Landau approach \cite{Chaikin}. The appropriate variables in the polymer case are the complete director field $\bf a$, describing the orientation and the degree of order, and the polymer density field $\rho$. Both fields are coupled by the continuity requirement (\ref{continuity}). The conservation of polymer mass will be satisfied globally. The phase transition will be controlled by the density (concentration) of the polymer as is the case for lyotropic nematic liquid crystals. 

Let us stress that for computational reasons all equations must remain regular also for vanishing order, i.e., they must be expressed by the full vector $\bf a$. Decomposition of the form ${\bf a} = a {\bf n}$, where $a$ is the degree of order, would result in a singularity of the form 0 times $\infty$ taking place in centers of defects, where $\nabla{\bf n}$ diverges while the degree of order vanishes. In contrast, $\bf a$ and its derivatives remain regular everywhere. Taking into account the definition (\ref{current}), Eq.\,(\ref{continuity}) is already of the correct form.

In the elastic free energy, instead of using the usual Frank terms for splay, twist, and bend of the director,
\begin{equation}
	f^{Frank} = {1\over 2}K_1 (\nabla\cdot{\bf n})^2 + {1\over 2}K_2 [{\bf n}\cdot(\nabla\times{\bf n})]^2 +
			{1\over 2}K_3 [{\bf n}\times(\nabla\times{\bf n})]^2, \\
			~ \label{frank}
\end{equation}
a new set of elastic terms must be used \cite{svensek-lozar,svensek-blanc}:
\begin{eqnarray}
	f^{el} &=& {1\over 2}L_1' (\partial_i a_j)^2 + {1\over 2}L_2' (\partial_i a_i)^2 \label{elastic}\\
			& & + {1\over 2}L_3' a_i a_j (\partial_i a_k)(\partial_j a_k)
			+ {1\over 2}L_4' (\epsilon_{ijk}a_k \partial_i a_j)^2.\nonumber
\end{eqnarray}
Unlike the Frank elastic parameters $K_i$, the elastic constants $L_i'$ do not depend on the degree of order.
To keep the number of elastic parameters at minimum, among all possible terms quadratic in the derivative we have retained only those non-vanishing in the limit of a fixed degree of order. 
Comparison of Eqs.\,(\ref{frank}) and (\ref{elastic}) in this limit relates the Frank constants $K_i$ to the constants $L_i$:
\begin{eqnarray}
	K_{1} & = & a^2 L_1' + a^2 L_2', \\
	K_{2} & = & a^2 L_1' + a^4 L_4', \\	
	K_{3} & = & a^2 L_1' + a^4 L_3'.
\end{eqnarray}
One observes that the dependence on the degree of order is different than in the case of the nematic tensor order parameter \cite{svensek-lozar,svensek-blanc}, where the corrections that break the degeneracy of splay and bend are cubic in the degree of order. Note that all three deformation modes (splay, twist, and bend) are degenerate in the limit of small degree of order, as indicated by the leading order dependence $K_i\propto a^2$. This would not be the case if one just used $\bf a$ in the Frank expression (\ref{frank}). Yet it must be so, because the direction $\bf n$ necessary to distinguish between splay, twist, and bend, is not defined as $a\to 0$. Generally, the Frank expression (\ref{frank}) can be used only perturbatively, i.e., for $|{\bf a}|\approx 1$, whereas the expression (\ref{elastic}) or similar can be used to describe the full range $0\le a\le 1$.

To establish a one-to-one correspondence between the two sets of elastic parameters, we make a further simplification of our model free energy expression (\ref{elastic}), while retaining full elastic anisotropy known to be significant in lyotropic liquid crystals. One of the possibilities is omitting the $L_4'$ term. Hence, the minimal free energy model in this case reads
\begin{eqnarray}
	f & = & {{1\over 2}}\rho C {\rho^*-\rho\over\rho^*+\rho}a^2 + {{1\over 4}}\rho C a^4\label{f_bulk} \\  
	  & + & {1\over 2}\rho L_1 (\partial_i a_j)^2 \label{f_elastic1} \\ 
	  & + & {1\over 2}\rho L_2 (\partial_i a_i)^2 
			+ {1\over 2}\rho L_3 a_i a_j (\partial_i a_k)(\partial_j a_k)\label{f_elastic23} \\ 
	  & + &	{{1\over 2}}G \left[\partial_i(\rho{a_i})-{{\rho^{\pm}}\over{\ell_0}}\right]^2 \label{f_G} \\ 
	  & + & {1\over 2}\chi (\rho-\rho_0)^2 \label{f_rho} \\ 
	  & + & {1\over 2}L_\rho (\partial_i\rho)^2, \label{f_gradrho}
\end{eqnarray}
where $C$ is a positive Landau constant describing the isotropic-nematic phase transition and $\rho^*$ is the transition density, 
\begin{eqnarray}
	L_1 & = & K_2/(\rho_0 a^2),\\
	L_2 & = & (K_1 - K_2)/(\rho_0 a^2),\\
	L_3 & = & (K_3 - K_2)/(\rho_0 a^4),
\end{eqnarray}
$\rho_0$ is the bulk equilibrium density, $\chi$ and $L_\rho$ are the density compressibility and the density variation correlation length. The nonlinear density factor in the first term of (\ref{f_bulk}) guarantees that the bulk nematic ordering stays limited to $|{\bf a}|<1$. In the terms (\ref{f_bulk})-(\ref{f_elastic23}) of the total free energy, corresponding to the ordering and elastic parts of the free energy, we have taken into account the fact that they need to be proportional to the number of molecules, i.e., to the local density.

The continuity requirement (\ref{continuity}) is taken into account by means of the penalty potential (\ref{f_G}) proportional to a coupling constant $G$ (the arbitrary length $\ell_0$ has been absorbed in $G$). For most of the time, the density of beginnings and ends of chains, $\rho^{\pm}$, will not be considered as a variable but as a fixed external parameter. 

What physical meaning can be attributed to the parameter $G$ that  defines the rigidity of the continuity constraint (\ref{continuity}), as this parameter does not exist if the continuity constraint (\ref{continuity}) is satisfied exactly? One notes that for the case $\rho^{\pm}  = 0$, there emerges a splay relaxation length scale $l_\rho = \sqrt{G/\chi}$ that controls the rigidity of the continuity constraint. Splay deformation of the director field at length scales much larger than $l_\rho$ does not bring about any substantial density variation, i.e., at this scale splay and density are decoupled. On the other hand, on length scales much shorter than $l_\rho$,  the coupling between splay and density becomes strong. This makes sense physically  as on large length scales the (weak) divergence of the polymer current is compensated by a spontaneous rearrangement of chain ends  -- which we are not taking into account -- while this is not effective at shorter length scales. It seems natural that $l_\rho$ should be in direct relation with the chain length \cite{Kamien}.

\subsection{Euler - Lagrange equations and their solution}

We will find the equilibrium configuration of both constitutive fields, i.e. the density field $\rho$ and the non-unit nematic director field $\bf a$, by minimizing the free energy at the constraint of global polymer mass conservation. The corresponding functional is thus
\begin{equation}
	F  =  \int\!\! {\rm d}V\, (f - \lambda \rho),\label{F-lambda}
\end{equation}
with the constraint
\begin{equation}
	\int\!\!{\rm d}V\, \rho = m_0 = {\rm const}, \label{mass_conservation}
\end{equation}
where $\lambda$ is a constant Lagrange multiplier. The minimization will be performed by following a quasi-dynamic evolution of the director and density fields of the form
\begin{eqnarray}
	\gamma{\partial a_i\over\partial t}  & = & \partial_j{\partial f\over\partial(\partial_j a_i)} - 
							{\partial f\over\partial a_i},\\
	\gamma{\partial \rho\over\partial t}  & = & \partial_j{\partial f\over\partial(\partial_j \rho)} - 
							{\partial f\over\partial \rho} + \lambda, \label{EL_rho}
\end{eqnarray}
where $\gamma$ is a formal parameter defining the time scale. Eq.\,(\ref{EL_rho}) shows that satisfying the constraint of mass conservation is easy: at every time step one subtracts from the density the homogeneous field 
\begin{equation}
	\Delta\rho = {1\over V}\int\!\!{\rm d}V\, \rho - m_0 \label{DeltaRho}
\end{equation}
so that the corrected density $\rho - \Delta\rho$ satisfies the mass conservation (\ref{mass_conservation}).

We use a tangentially degenerate boundary condition for the director, i.e., the director is everywhere parallel to the surface of the sphere, while it is allowed to rotate freely in the tangential plane. For the density we use free boundary condition, i.e., the normal (radial) component of the density gradient is zero. 

The initial condition is $\rho({\bf r})=\rho_0$ for the density field and ${{\bf a}({\bf r})}=0$ plus a small random perturbation for the director field. The equations are solved by an open source finite volume solver (OpenFOAM accessible at http://www.openfoam.com/) on a 50 x 50 x 50 cubic mesh (size of the box containing the sphere), which is deformed and refined near the surface of the sphere to define a smooth boundary.

\section{Results}

In what follows we present the steady state solutions of the quasi-dynamic evolution of the director and density fields corresponding to direct solutions of the Euler-Lagrange equation. The steady-state solutions obviously satisfy 
\begin{eqnarray}
	0  & = & \partial_j{\partial f\over\partial(\partial_j a_i)} - 
							{\partial f\over\partial a_i},\\
	0  & = & \partial_j{\partial f\over\partial(\partial_j \rho)} - 
							{\partial f\over\partial \rho} + \lambda,
\end{eqnarray}
with the appropriate mass constraint (\ref{mass_conservation}). Unless stated otherwise, in the numerical simulation we take the following dimensionless values for the elastic constants entering our model: $A=C=1$, $L_1=1$, $L_2=L_3=0$, $G=1$, $\chi=1$, $L_\rho=1$. This means that in these units the nematic correlation length (the characteristic size of the defect core) is 1, as is the characteristic length $l_\rho$ of the director-density coupling. The bulk equilibrium density is set to $\rho_0 = 1.5$ and the nematic transition threshold density to $\rho^*=0.5$. The only remaining parameter, $\ell_0$, merely scales the source density $\rho^\pm$, Eq.\,(\ref{f_G}), and can be set to $\ell_0=1$ without loss of generality.

The tangential boundary condition for the director implies a disclination of the total strength +2 on the surface of the sphere \cite{Chaikin}. Most usually this means a pair of disclinations with strength +1 each.  Quite generally, due to the expensive splay deformation in nematic polymers, circular disclinations are preferred to the radial ones which would be favoured in a regular nematic because they are compatible with a cheaper three-dimensional configuration.

\begin{figure}
\begin{center}
	\mbox{\subfigure[~$\bar{f}=-0.07152$]{\includegraphics[width=42mm]{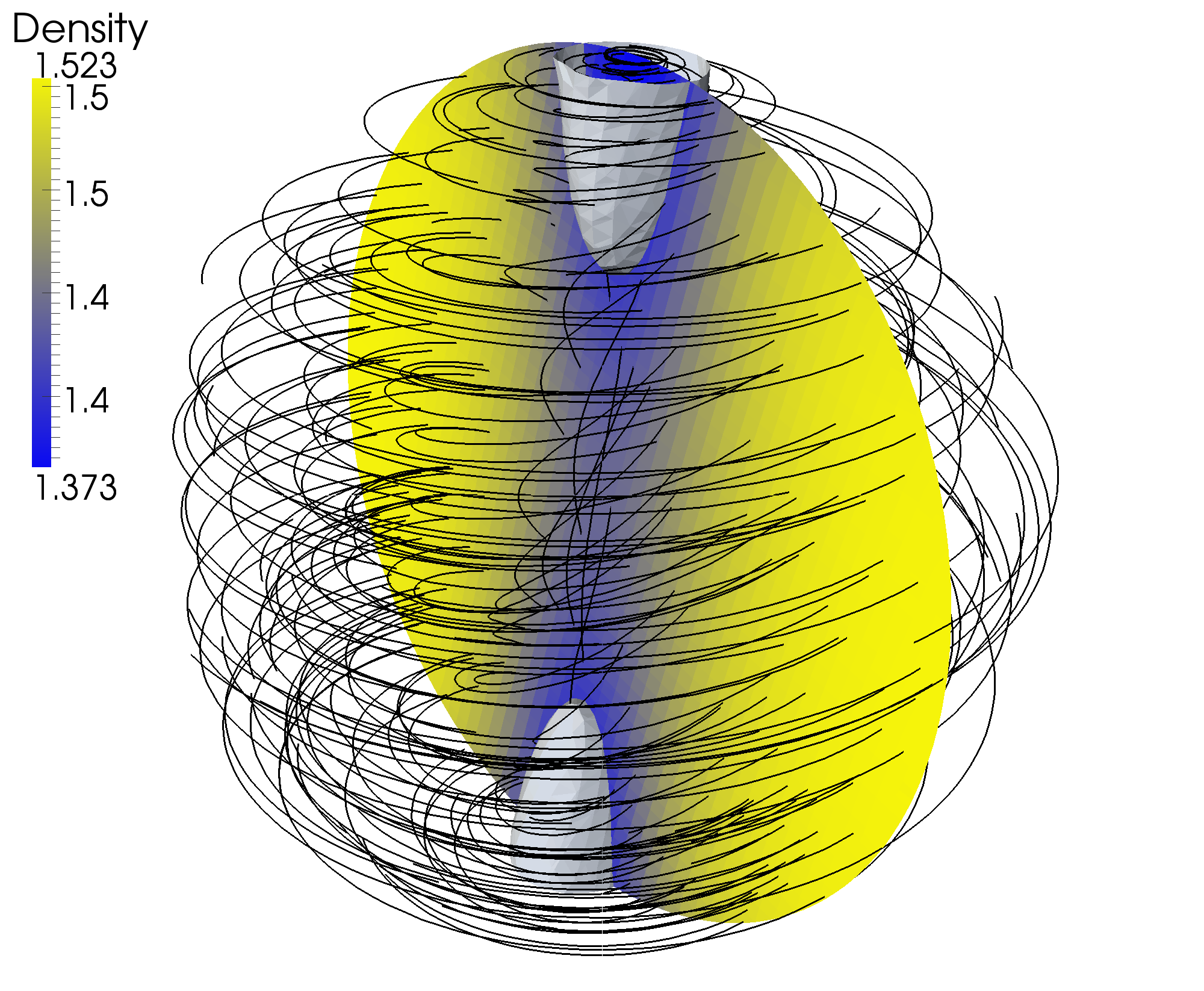}}\hspace{0cm}
		  \subfigure[~$\bar{f}=-0.08966$]{\includegraphics[width=42mm]{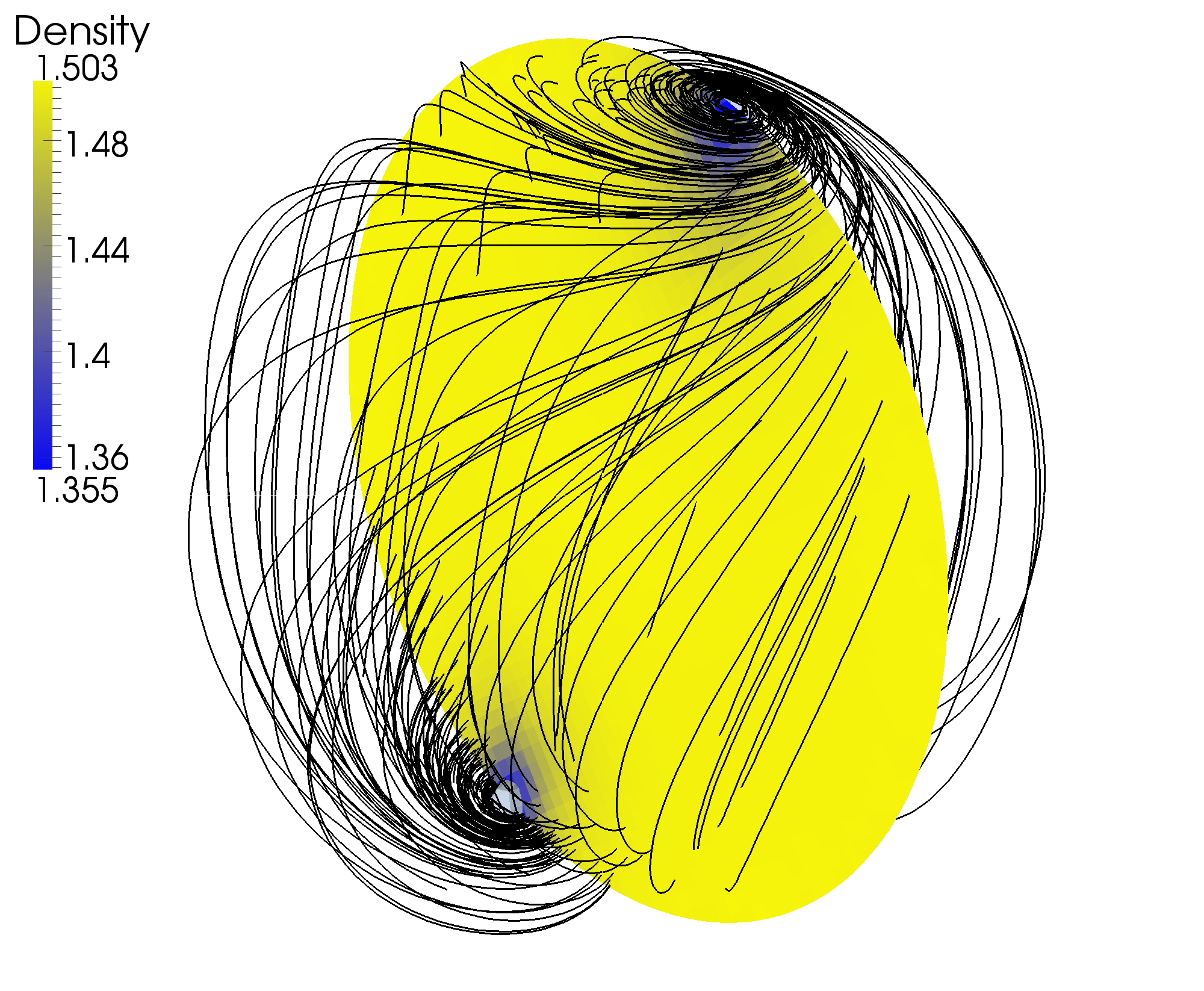}}\hspace{0cm}  }
\caption{The director and density fields in the case of no sources, $\rho^{\pm} = 0$. The radius of the confining sphere is (a) 10 and (b) 32 in units of the nematic correlation length. The average free energy density $\bar{f}$ of the configurations is indicated; it should be compared for systems of the same size.  The director field is shown by random tracer paths, on the iso-contour the value of the nematic order is half the bulk equilibrium value. The density profile is shown in the cutting plane.}
\label{no-sources}
\end{center}
\end{figure}

\begin{figure}
\begin{center}
	\mbox{\subfigure[~$\bar{f}=-0.07093$]{\includegraphics[width=42mm]{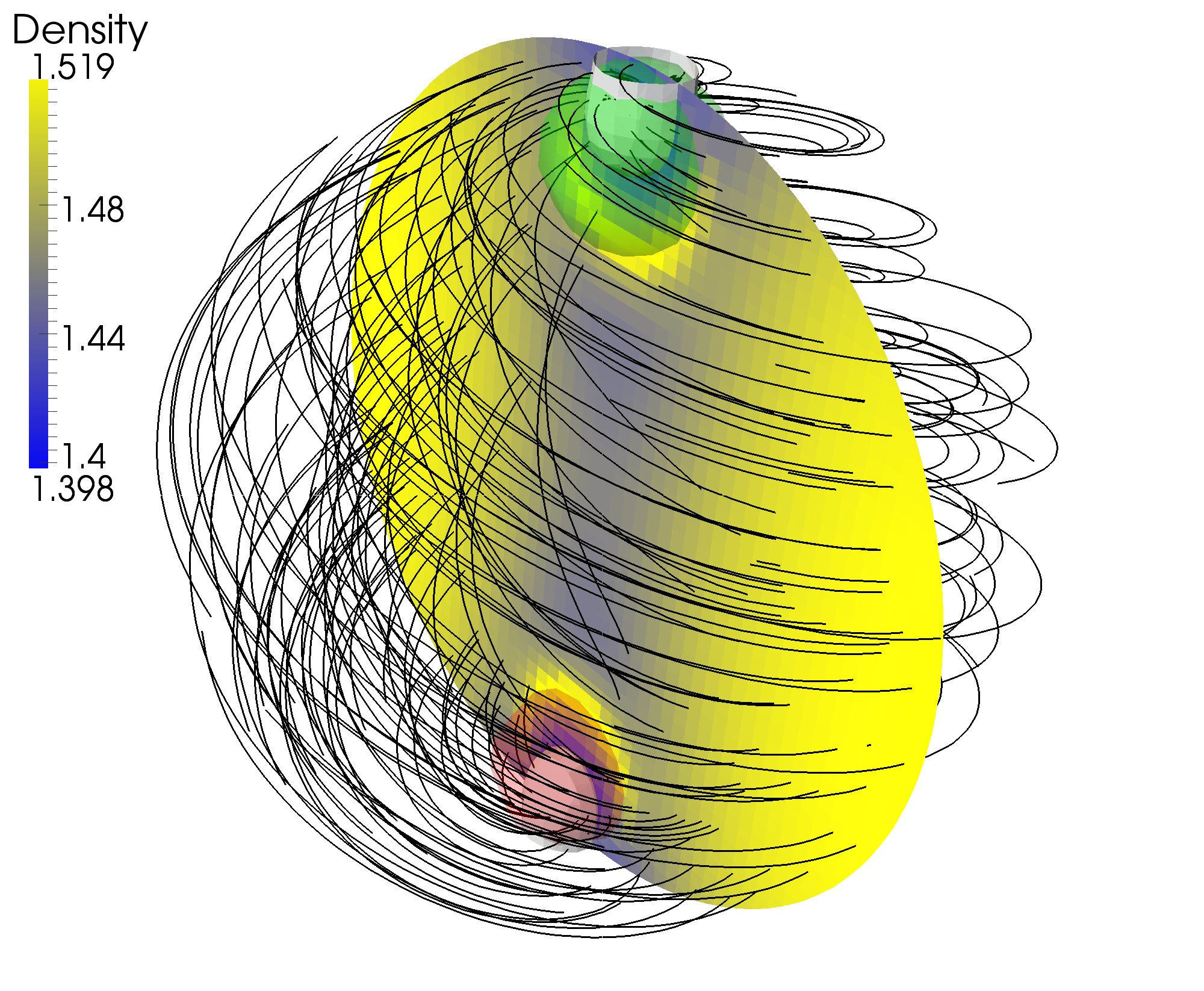}}\hspace{0cm}
		  \subfigure[~$\bar{f}=-0.08848$]{\includegraphics[width=42mm]{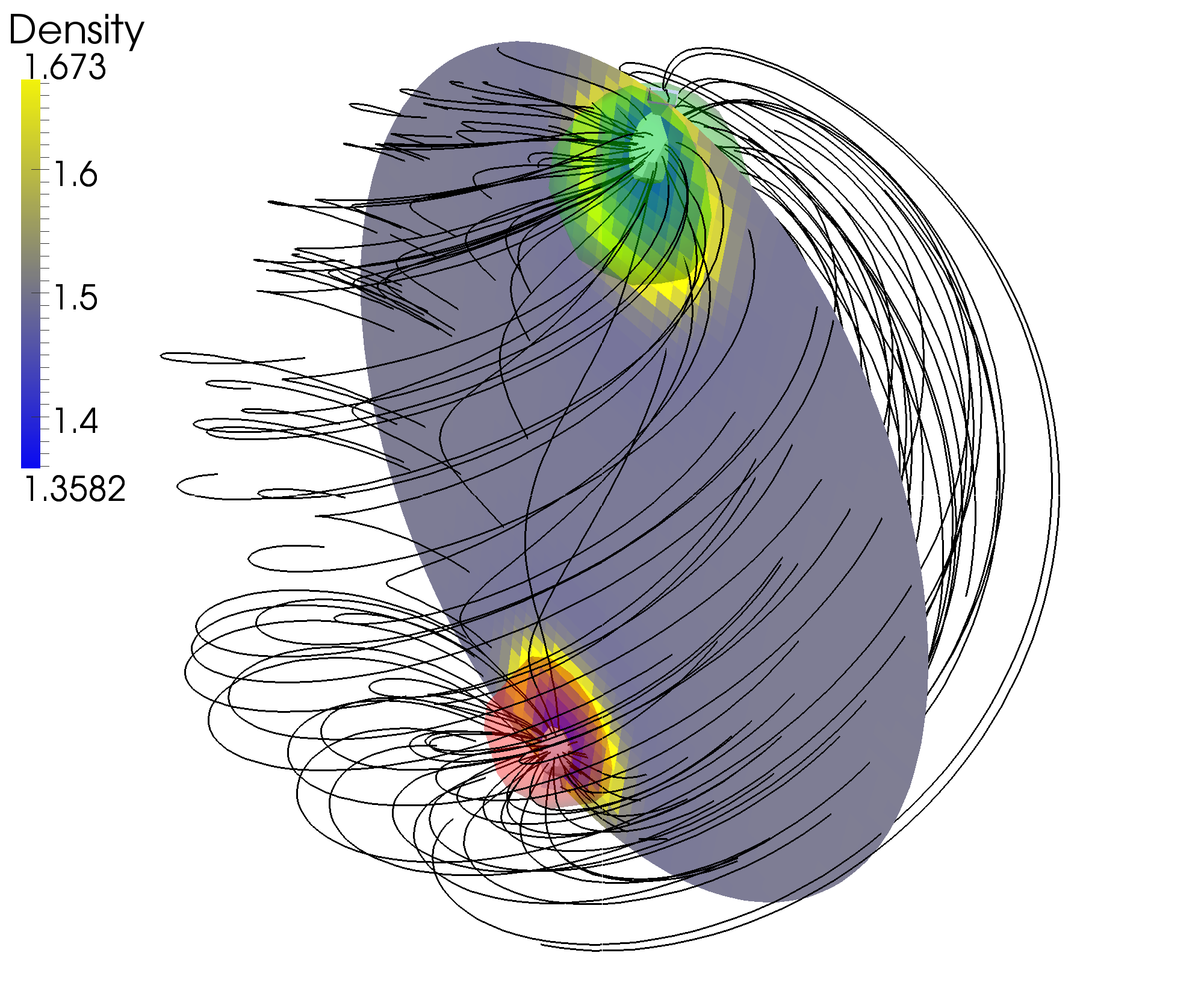}}\hspace{0cm}  }

\caption{Dipolar sources: $\rho^{\pm}=1$ in the green region, $\rho^{\pm}=-1$ in the red region.  
The radius of the confining sphere is (a) 10 and (b) 32. 
The key is explained in Fig.\,\ref{no-sources}.}
\label{dipolar-sources}
\end{center}
\end{figure}

\subsection{No sources}

We first of all find the solutions for the case of no sources, $\rho^{\pm} = 0$, in a spherical hard enclosure. Obviously, see Fig.\,\ref{no-sources}(a), the geometry of the director field is pronouncedly toroidal of an inverse spool type (note that the director field is represented by trajectories of random tracers), while the density field shows a moderate depletion on the cylindrical axis. Most notably, the density depleted axial core is only partially orientationally disordered, i.e. the disclination line is avoided by the director spiraling becoming increasingly vertical in the core. The iso-contours of the nematic ordering ($|{\bf a}|$) near the poles show that the defects are predominantly point-like but not well isolated due to the small system size.

The equilibrium profiles of the density and director field depend crucially on the size of the confining sphere. Making the sphere bigger, see Fig.\,\ref{no-sources}(b), leads to a pronounced localization of the axial density depletion range to polar regions, while the rest of the spherical volume is almost uniformly packed with the polymer as the director configuration is more spiral and less toroidal if compared to the case of tighter confinement. This can only come about at the cost of a deformed director field that develops two opposite vortices (with strength +1 each) in the polar regions. 

\subsection{Fixed sources}

Introduction of sources further modifies the constitutive fields. We first of all take two oppositely "charged" localized sources, $\rho^{\pm}=1$ and $\rho^{\pm}=-1$ located symmetrically at the two poles, see Fig.\,\ref{dipolar-sources}. Since the sources correspond to oppositely ''charged'' ends of the chain we can refer to this configuration as a {\sl dipolar} configuration. The introduction of the two sources has only a limited effect on the configuration, as the configurations in Fig.\,\ref{no-sources} already exhibit a cylindrical symmetry (with the distinction that there the direction of the symmetry axis is selected spontaneously). The major difference is that now the two surface defects become more radial, and hence the configuration is less toroidal. This is expected since now the sources allow more splay around the defects. The pulse-like behaviour of the density around the point defects is a signature of the step-like profile of the source densities.

\begin{figure}[t!]
\begin{center}
	\mbox{\subfigure[~$\bar{f}=-0.07114$]{\includegraphics[width=42mm]{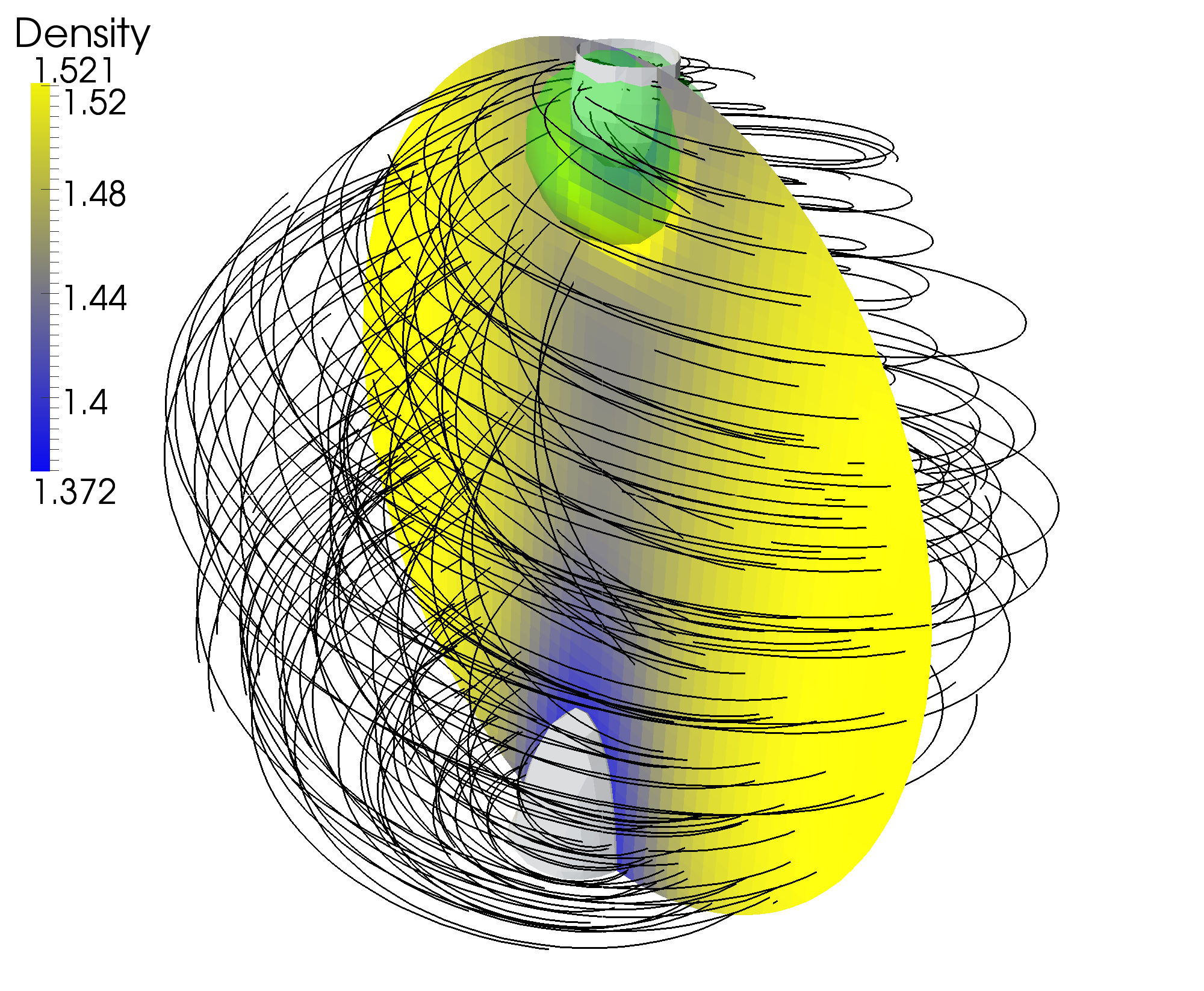}}\hspace{0cm}
		  \subfigure[~$\bar{f}=-0.08958$]{\includegraphics[width=42mm]{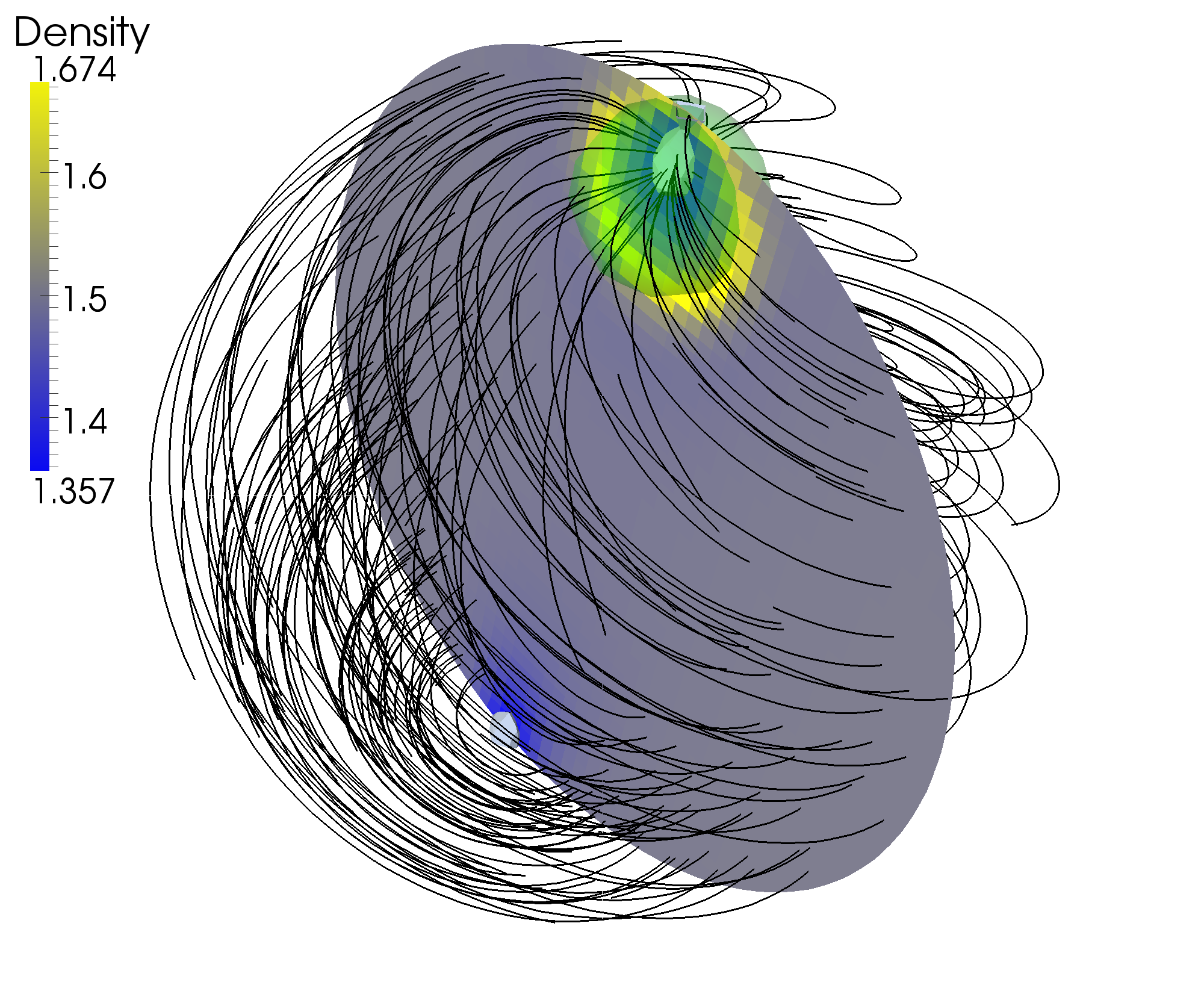}}\hspace{0cm}  }
\caption{Single source, $\rho^{\pm}=1$ in the green region, a corresponding negative $\rho^\pm$ is evenly distributed elsewhere so that the total source is zero. The radius of the confining sphere is (a) 10 and (b) 32. The key is explained in Fig.\,\ref{no-sources}.}
\label{single-source}
\end{center}
\end{figure}

For completeness we also show, see Fig.\,\ref{single-source}, the case of a configuration characterized by a single localized source, while a source density of opposite sign is evenly distributed over the whole system such that the total source is zero. This would correspond, e.g., to the polymer chain with one end fixed to the wall and the other end free --a situation often encountered while DNA is being packed into a capsid. While the single source configurations are quite similar to the dipolar case, it is instructive to observe the difference between the two defects of the single source configurations. The defect located at the end opposite to the source is almost entirely circular, in contrast to the defect located at the source. 

\begin{figure}[t!]
\begin{center}
	\mbox{\subfigure[~$\bar{f}=-0.06910$]{\includegraphics[width=42mm]{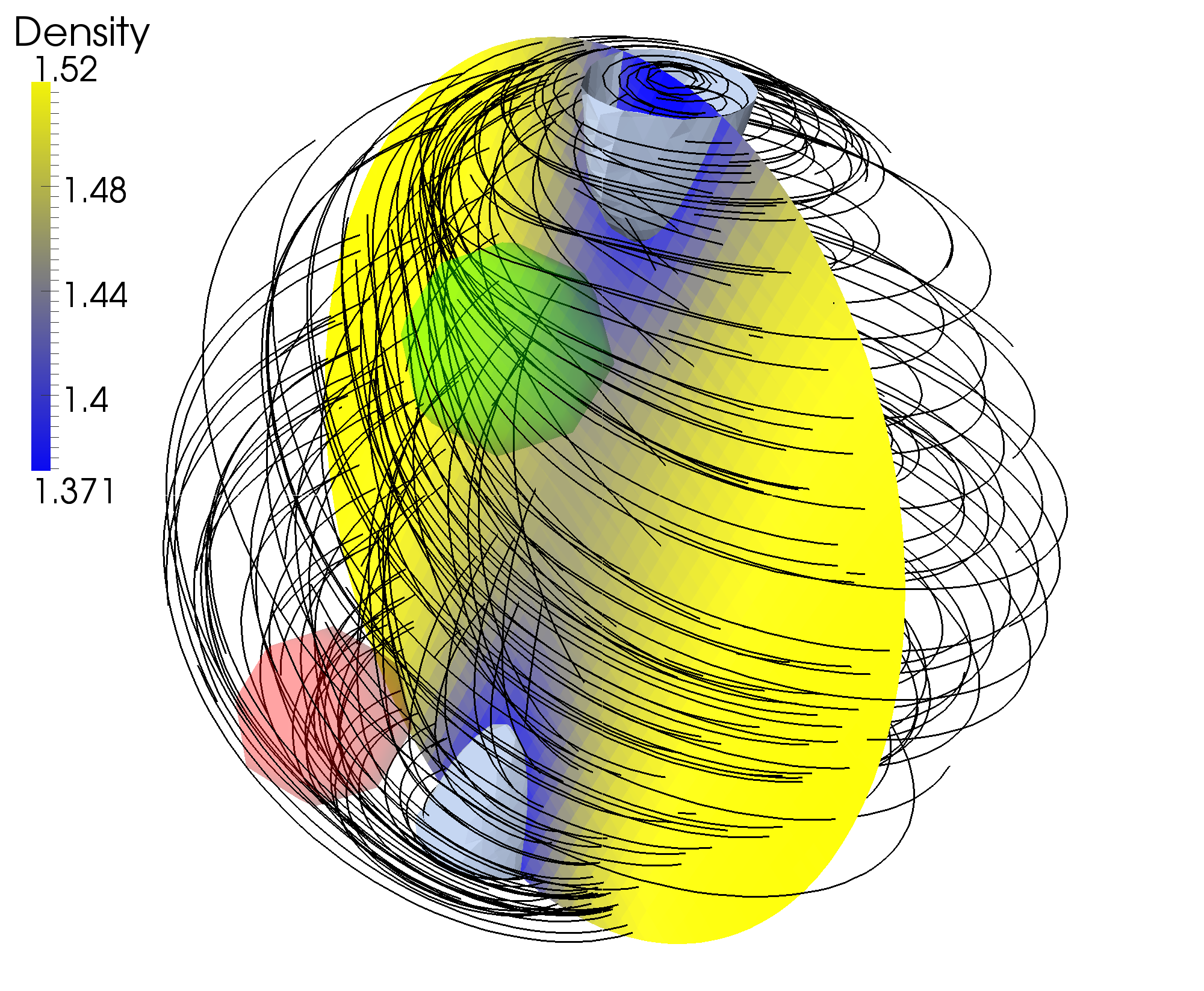}}\hspace{0cm}
		  \subfigure[~$\bar{f}=-0.08807$]{\includegraphics[width=42mm]{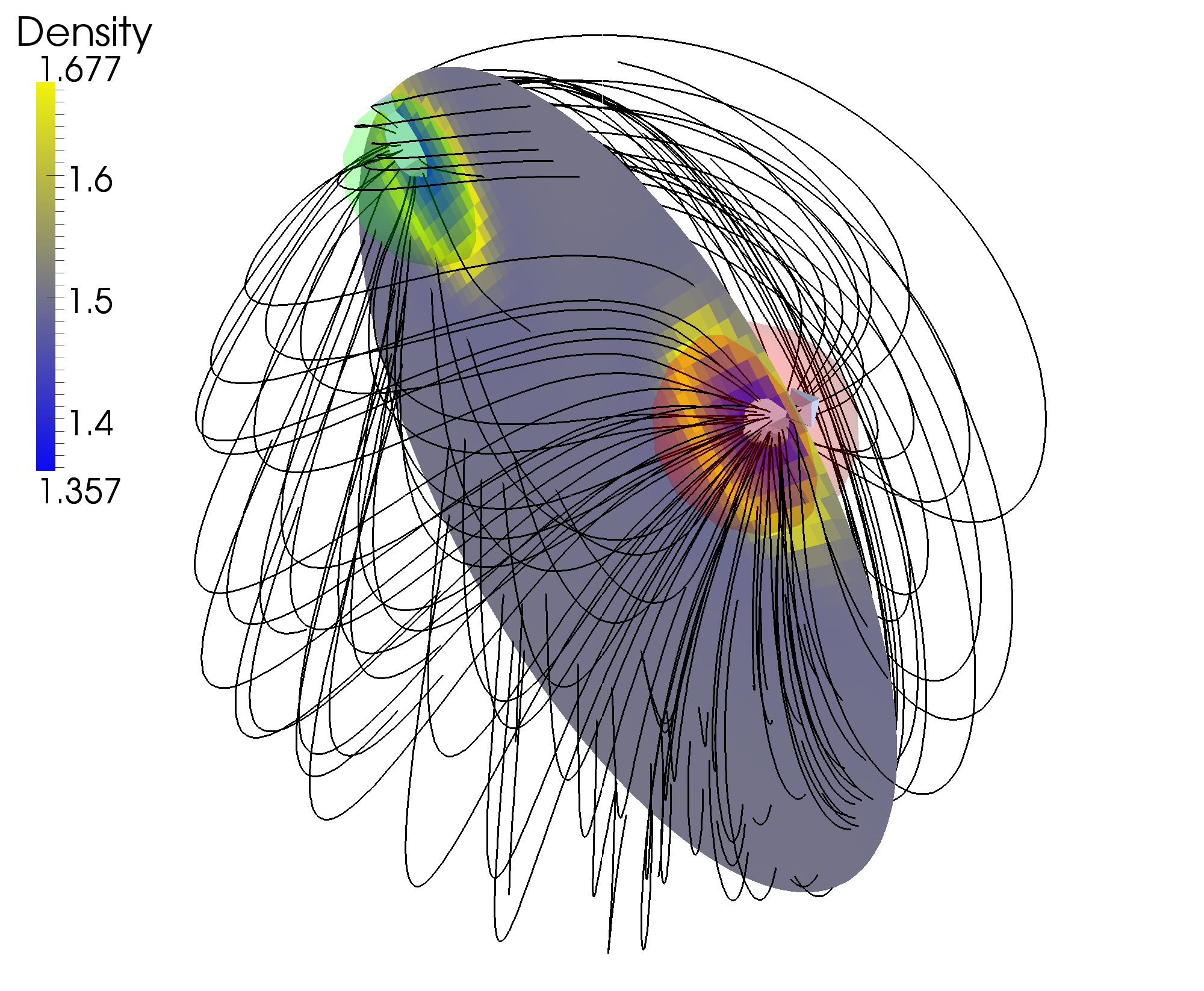}}\hspace{0cm}  }
\caption{A $\rho^{\pm}=1$ (green) and $\rho^{\pm}=-1$ (red) source placed on the $z$ and $x$ axes. The radius of the confining sphere is (a) 10 and (b) 32. The key is explained in Fig.\,\ref{no-sources}.}
\label{oblique-sources}
\end{center}
\end{figure}

An oblique placement of the localized $\rho^\pm=1$ and $\rho^\pm=-1$ sources, Fig.\,\ref{oblique-sources}, shows a particularly significant dependence on the size of the confining sphere. In the case of strong confinement, corresponding to a smaller radius of the sphere, the configuration is almost unaffected by the presence of the sources. While in the case of weaker confinement, corresponding to a larger radius of the sphere, the director configuration is completely governed by the two sources and shows the same symmetry as the positions of the sources.

\subsection{Compressibility and density correlation length effects}

\begin{figure}
\begin{center}
	\mbox{\subfigure[~$\bar{f}=-0.07631$]{\includegraphics[width=42mm]{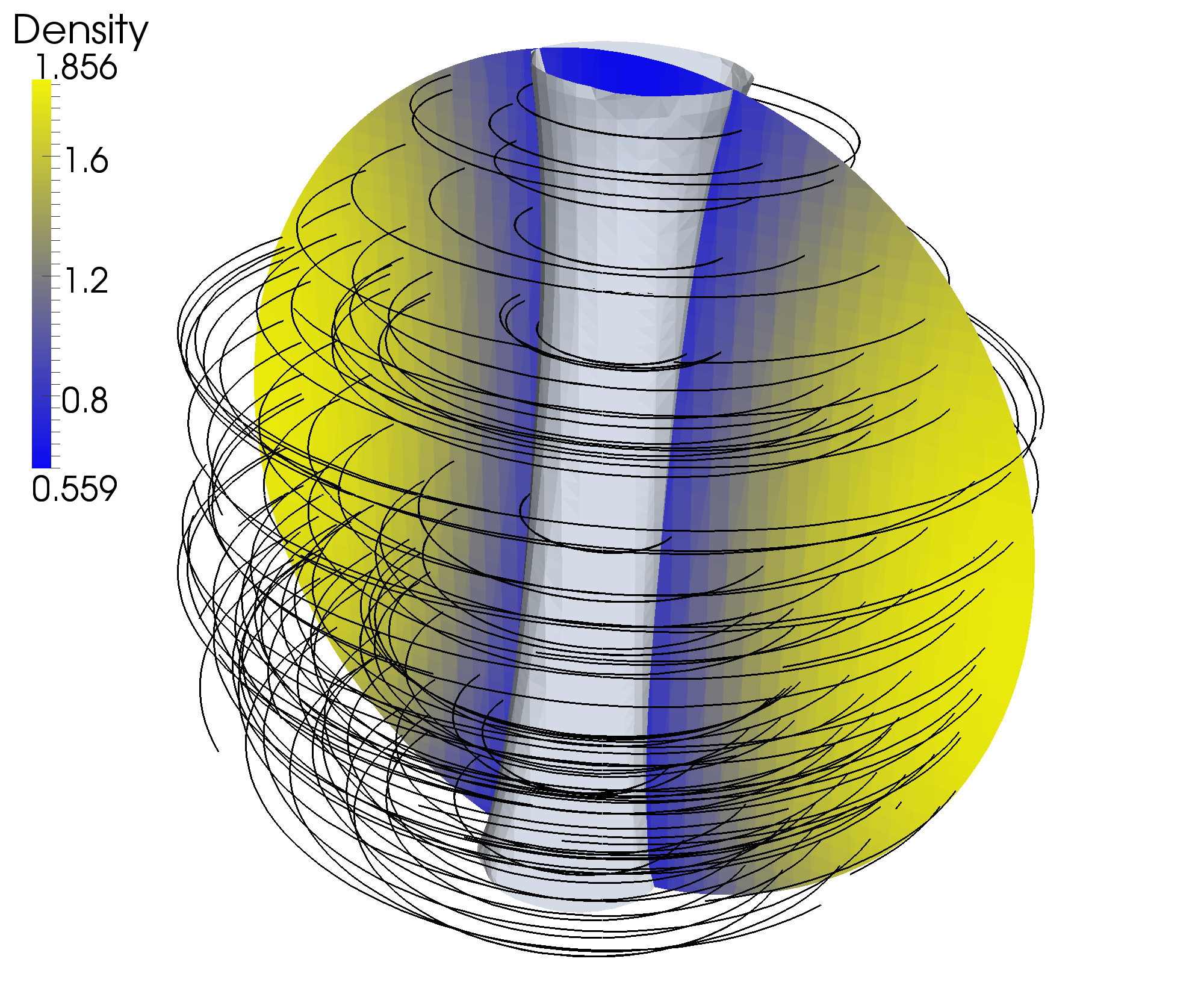}}\hspace{0cm}
		  \subfigure[~$\bar{f}=-0.09272$]{\includegraphics[width=42mm]{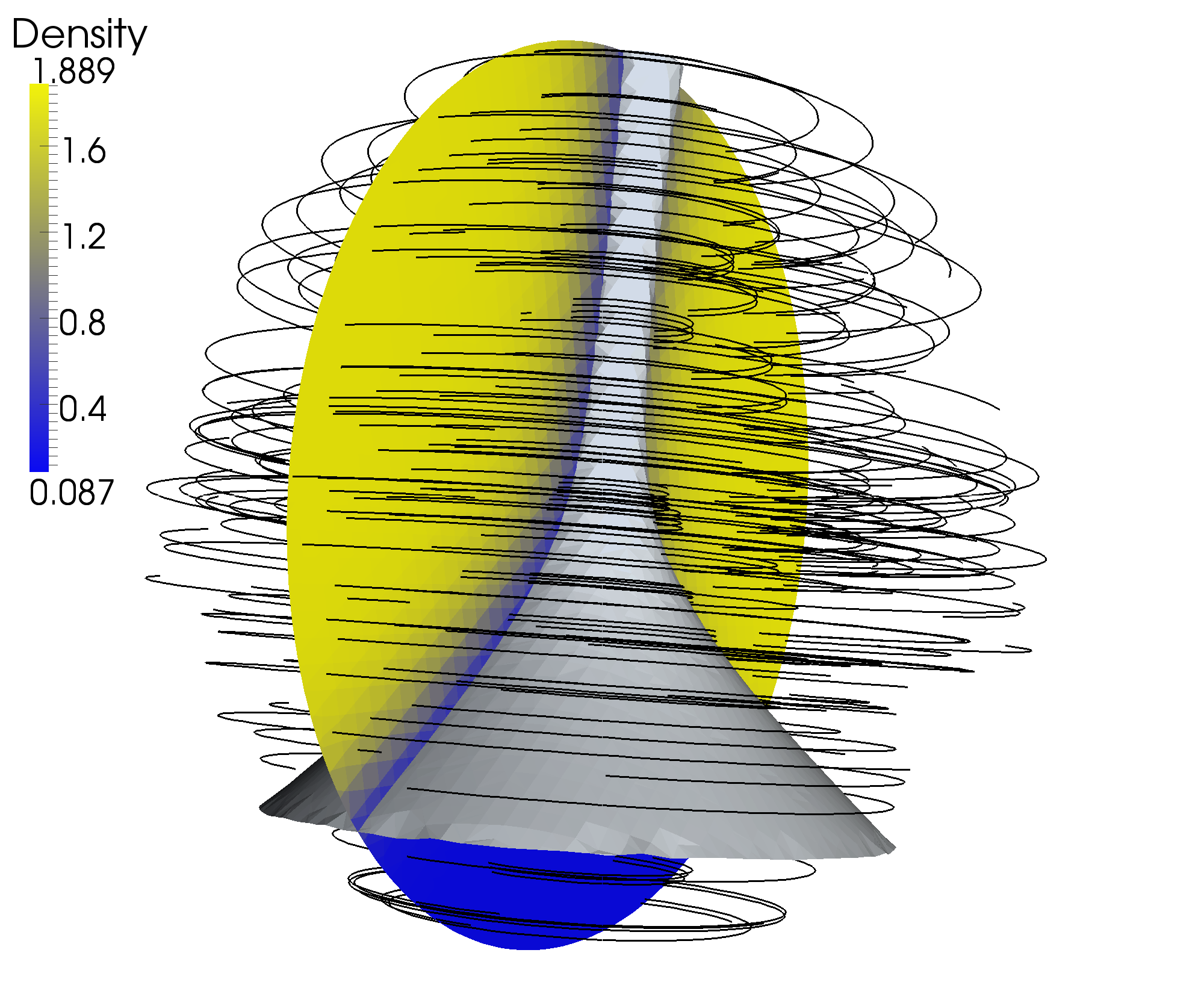}}\hspace{0cm}  }
\caption{Density variation cheaper, no sources. The radius of the confining sphere is (a) 10 and (b) 32. The key is explained in Fig.\,\ref{no-sources}.}
\label{soft-no-sources}
\end{center}
\end{figure}

The director configurations and packing density change significantly if we make density changes, as quantified by the compressibility modulus $\chi$ and the density correlation length $L_\rho$, less costly in energy terms. To demonstrate this we choose rather drastically $\chi=0.1$ and $L_\rho=0.33$, which nevertheless still yields $\rho({\bf r})>0$ everywhere within the enclosed volume. The severely confined configuration with no sources, 
Fig.\,\ref{soft-no-sources}(a), is in this case purely toroidal and exhibits a clear axial disclination line, while the density field shows a strong depletion around the axis of cylindrical symmetry. Moreover, in the case of a bigger confining sphere, Fig.\,\ref{soft-no-sources}(b), the polymer is completely depleted from a part of the available space, breaking the polar symmetry. It appears that stronger confinement operating for smaller sphere radii induce a toroidal packing irrespective of all the other terms in the free energy. This finding should be particularly relevant for the DNA packing within viral capsids and points to a {\sl universality} of the inverse spool configuration.

\begin{figure}
\begin{center}
	\mbox{\subfigure[~$\bar{f}=-0.07308$]{\includegraphics[width=42mm]{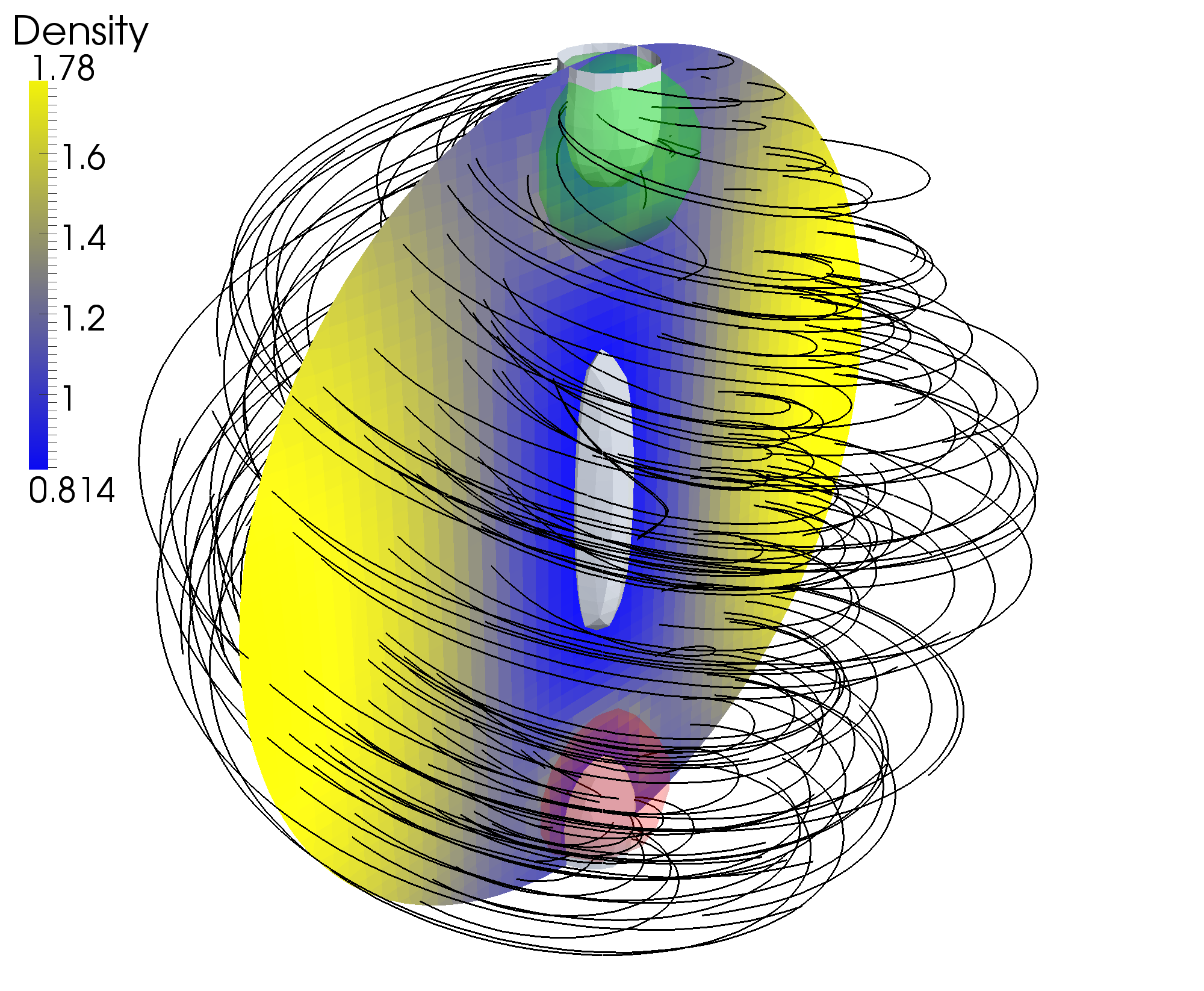}}\hspace{0cm}
		  \subfigure[~$\bar{f}=-0.08925$]{\includegraphics[width=42mm]{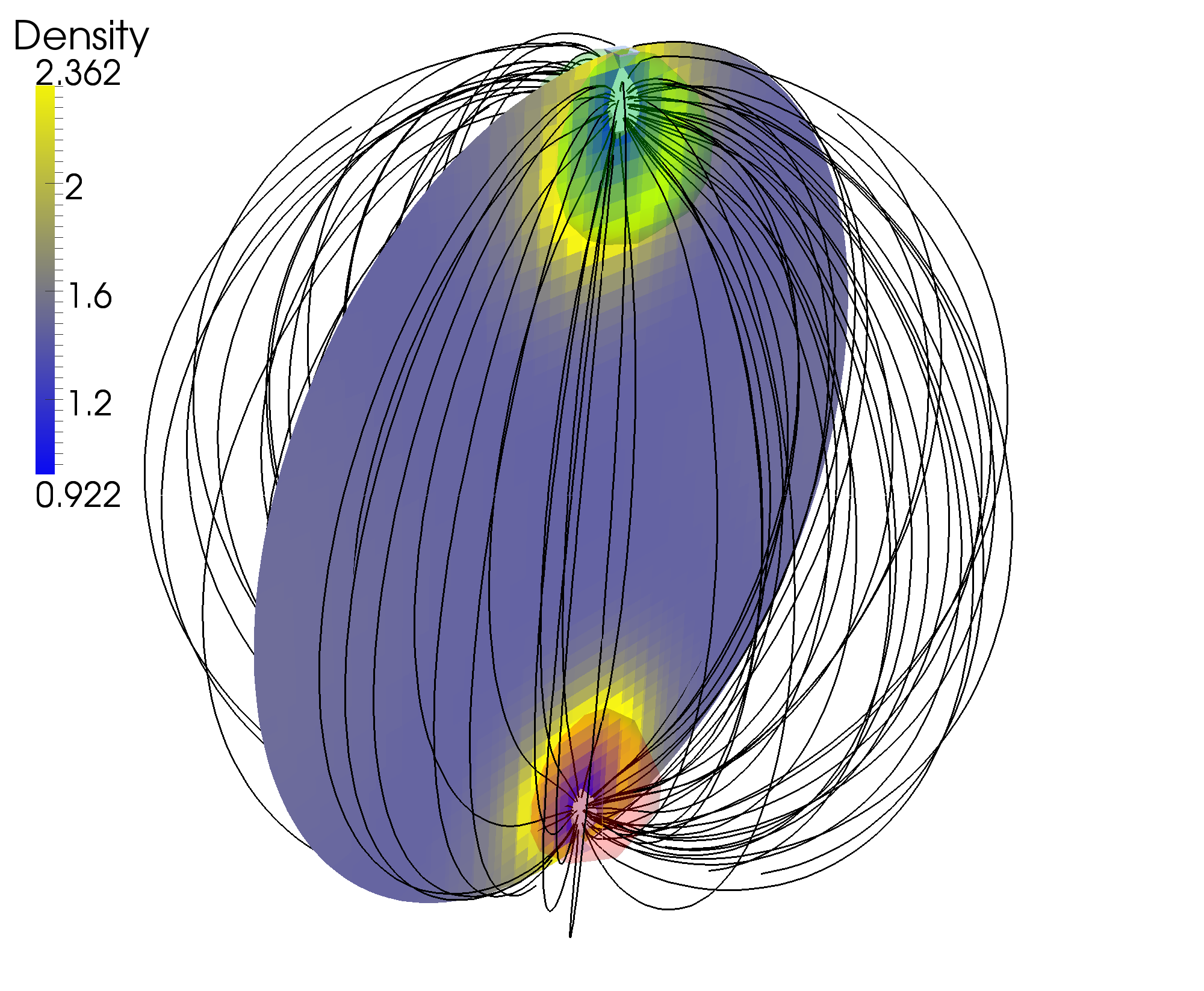}}\hspace{0cm}  }
\caption{Density variation cheaper, dipolar sources as in Fig.\,\ref{dipolar-sources}. The radius of the confining sphere is (a) 10 and (b) 32. The key is explained in Fig.\,\ref{no-sources}.}
\label{soft-dipolar-sources}
\end{center}
\end{figure}

The configurations with the dipolar sources show a pronounced central density depletion region and a stronger nematic order reduction in case of severe confinement corresponding to smaller sphere radii, Fig.\,\ref{soft-dipolar-sources}(a). The case of a bigger sphere radii again leads to purely polar defects, Fig.\,\ref{soft-dipolar-sources}(b), with a density variation showing a local spherical symmetry. In regular nematics, the latter director configuration is readily encountered and is known as the {\sl bipolar director configuration} \cite{zumer-crawford}. 

\begin{figure}
\begin{center}
	\mbox{\subfigure[~$\bar{f}=-0.07458$]{\includegraphics[width=42mm]{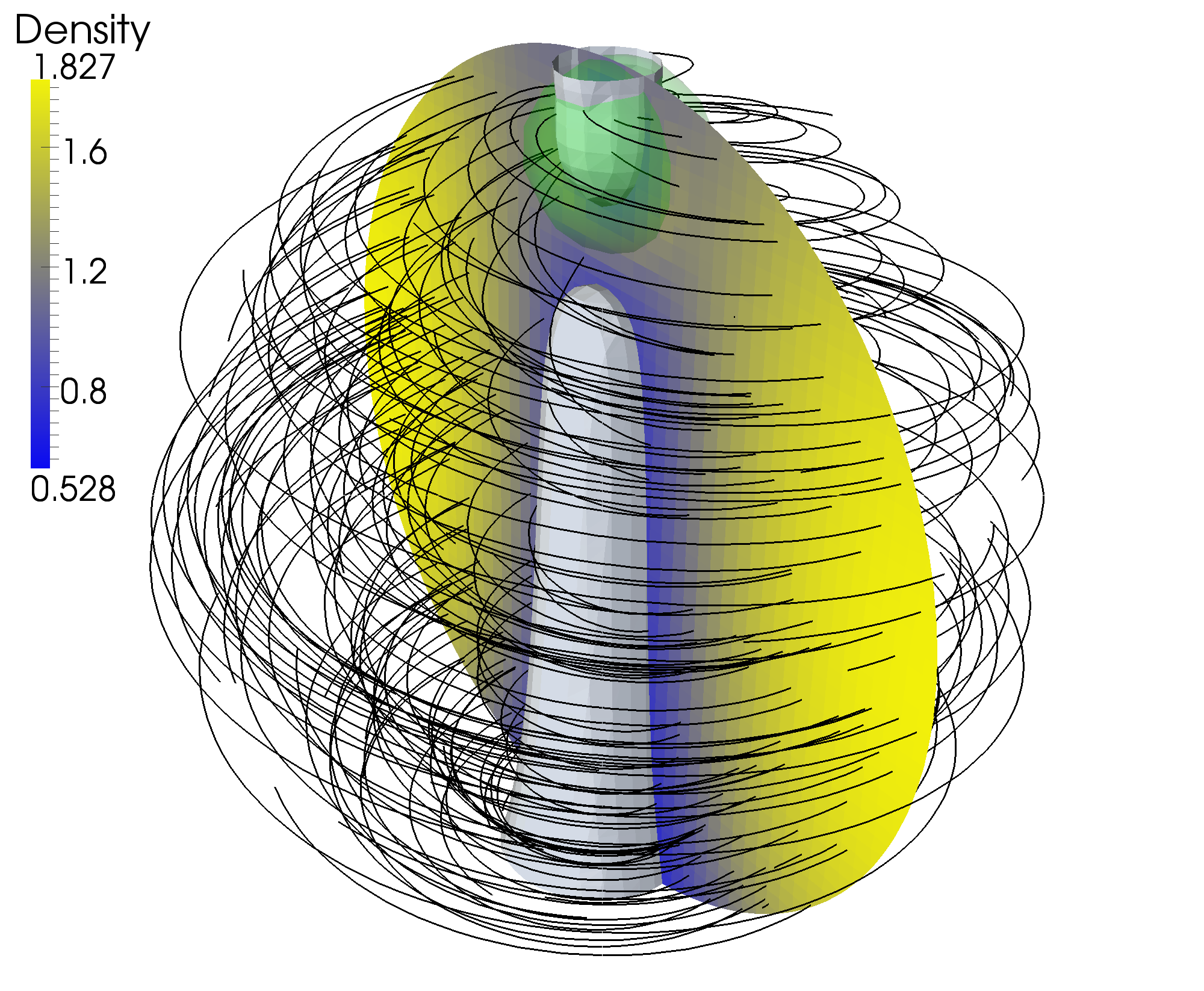}}\hspace{0cm}
		  \subfigure[~$\bar{f}=-0.09182$]{\includegraphics[width=42mm]{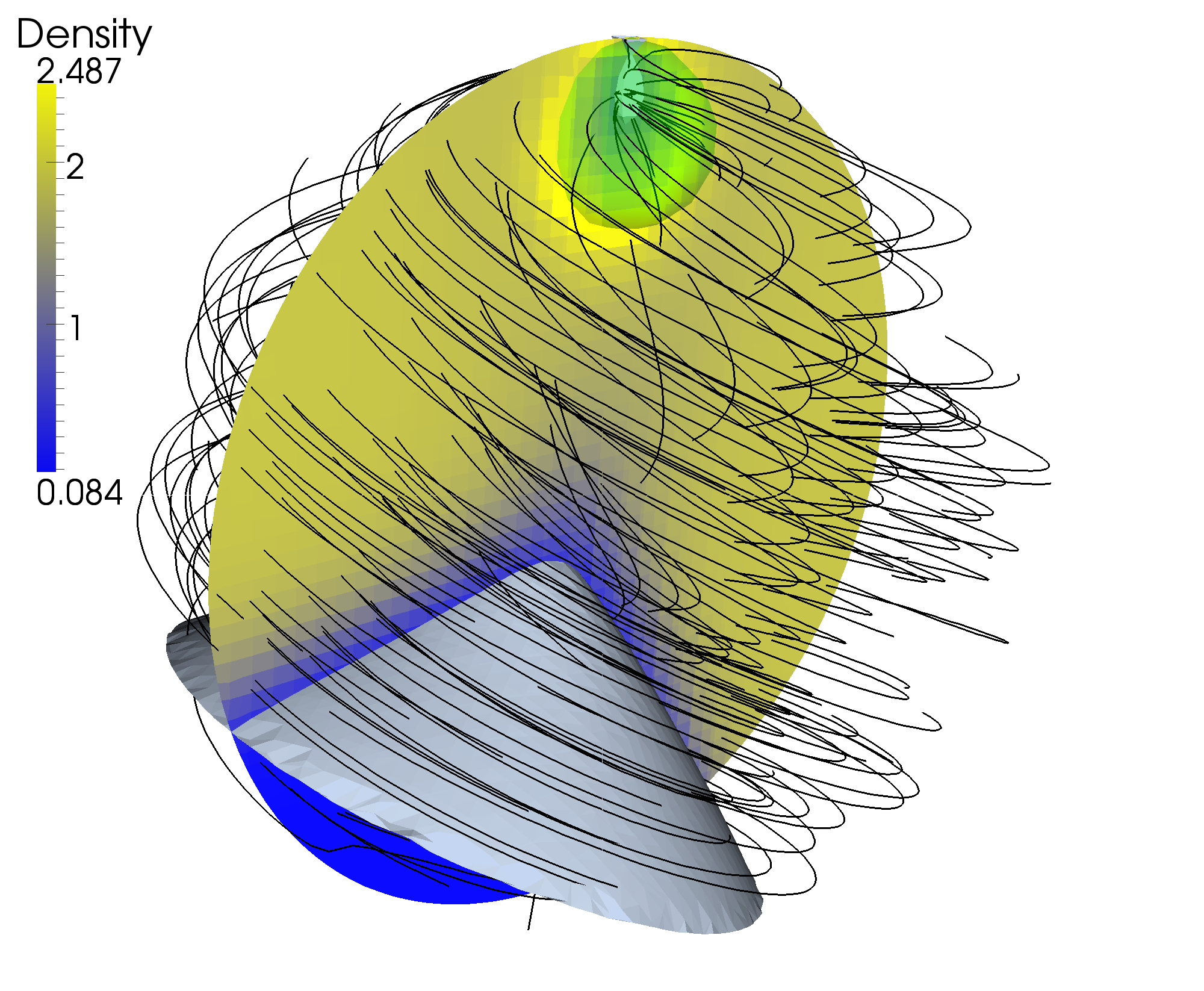}}\hspace{0cm}  }
\caption{Density variation cheaper, single source as in Fig.\,\ref{single-source}. The radius of the confining sphere is (a) 10 and (b) 32. The key is explained in Fig.\,\ref{no-sources}.}
\label{soft-single-source}
\end{center}
\end{figure}

\begin{figure}
\begin{center}
	\mbox{\subfigure[~$\bar{f}=-0.07374$]{\includegraphics[width=42mm]{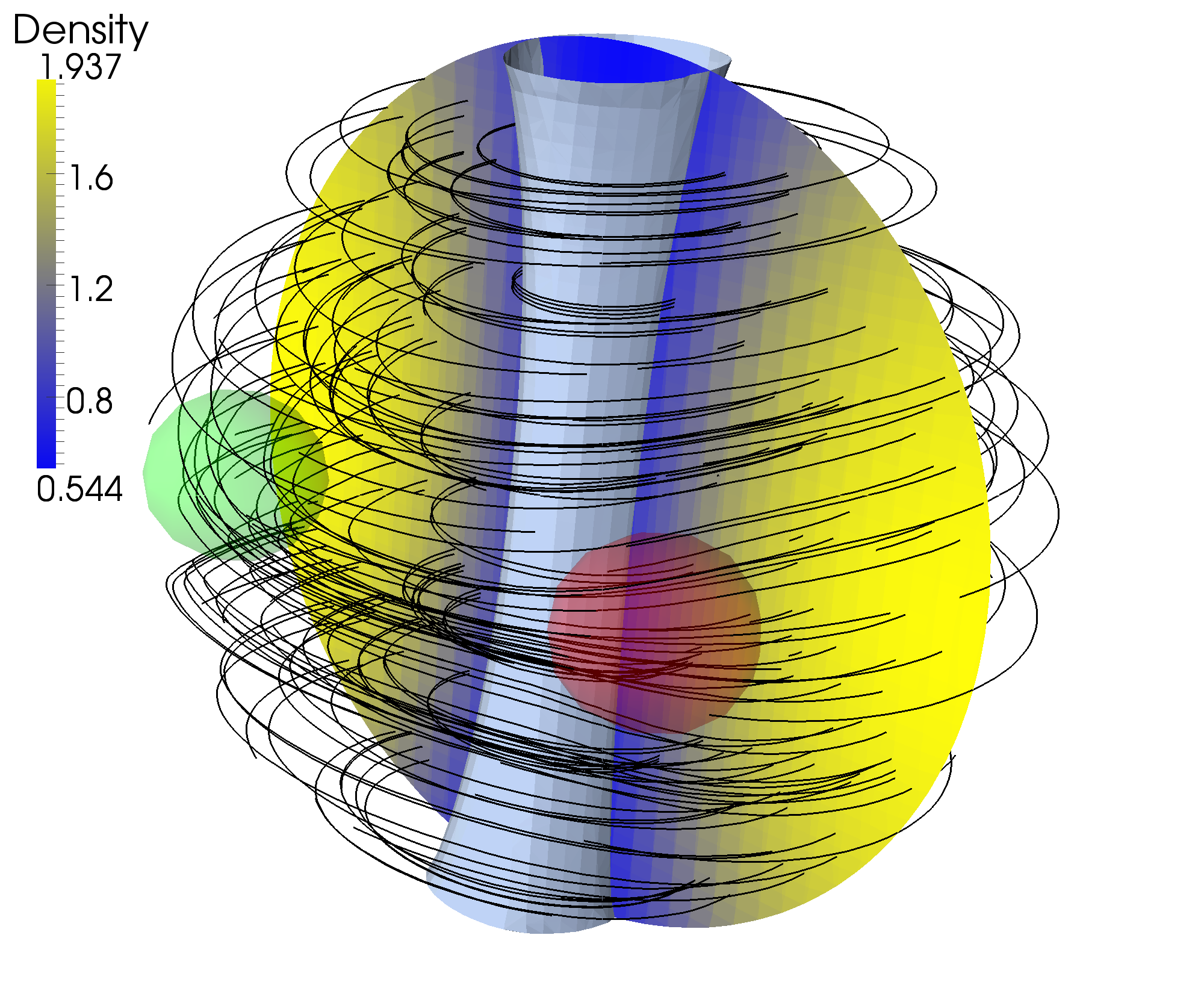}}\hspace{0cm}
		  \subfigure[~$\bar{f}=-0.08988$]{\includegraphics[width=42mm]{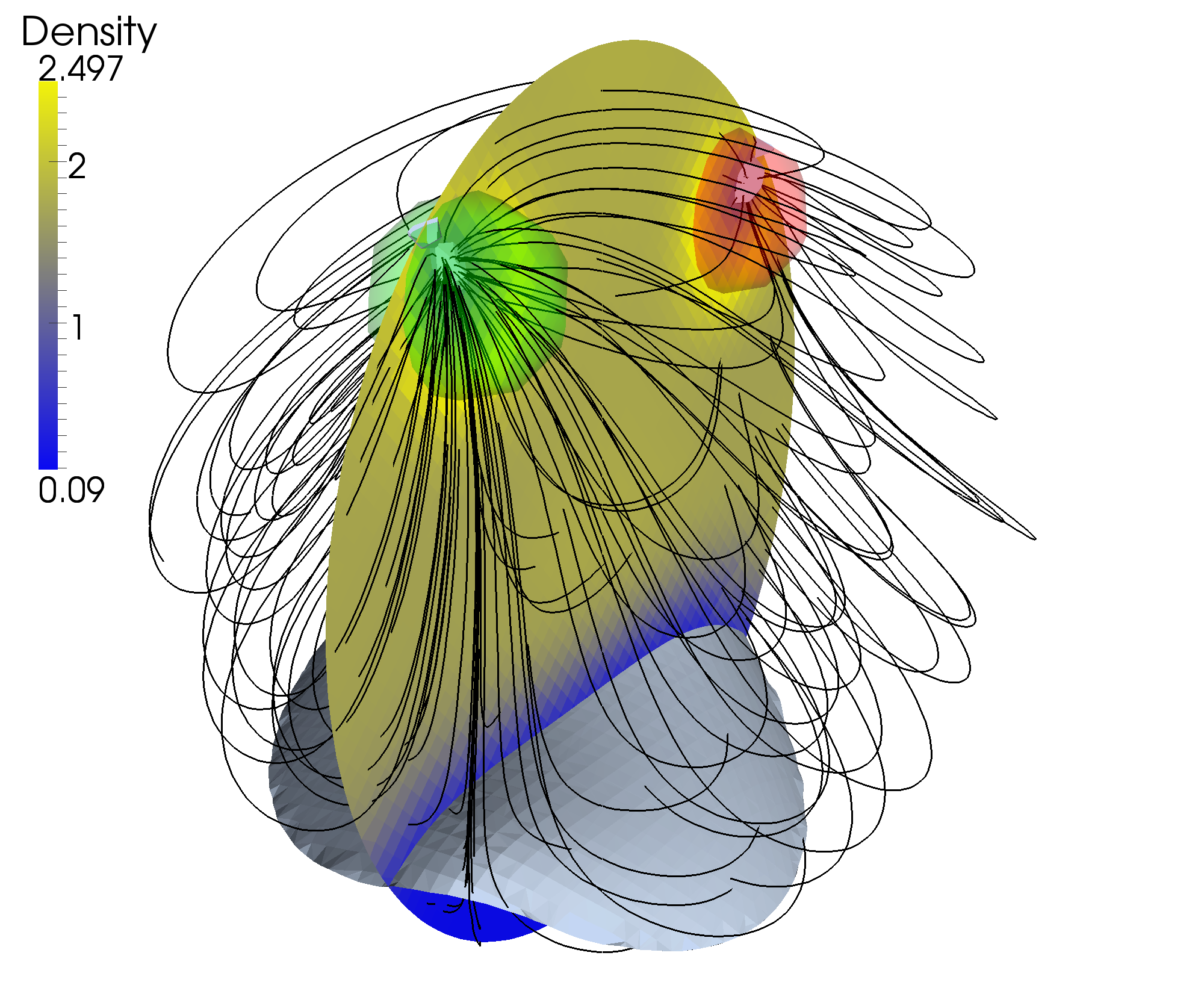}}\hspace{0cm}  }
\caption{Density variation cheaper, sources placed on the $z$ and $x$ axes as in Fig.\,\ref{oblique-sources}. The radius of the confining sphere is (a) 10 and (b) 32. The key is explained in Fig.\,\ref{no-sources}.}
\label{soft-oblique-sources}
\end{center}
\end{figure}

The single source cases, Fig.\,\ref{soft-single-source}, again show a strong axial polymer depletion and decrease of ordering (a), as well as almost complete depletion around a single pole within the sphere (b). The configurations with oblique sources, Fig.\,\ref{soft-oblique-sources} show similar features as before, plus a complete depletion in the part of the bigger sphere which is somewhat different to Figs.\,\ref{soft-no-sources} and \ref{soft-single-source}. In the latter cases the depletion conveniently removed one of the defects, whereas in the present case it is the most sluggish part of the director field which is depleted.

\subsection{Multiple sources}

\begin{figure}
\begin{center}
	\mbox{\subfigure[~$\bar{f}=-0.08740$]{\includegraphics[width=42mm]{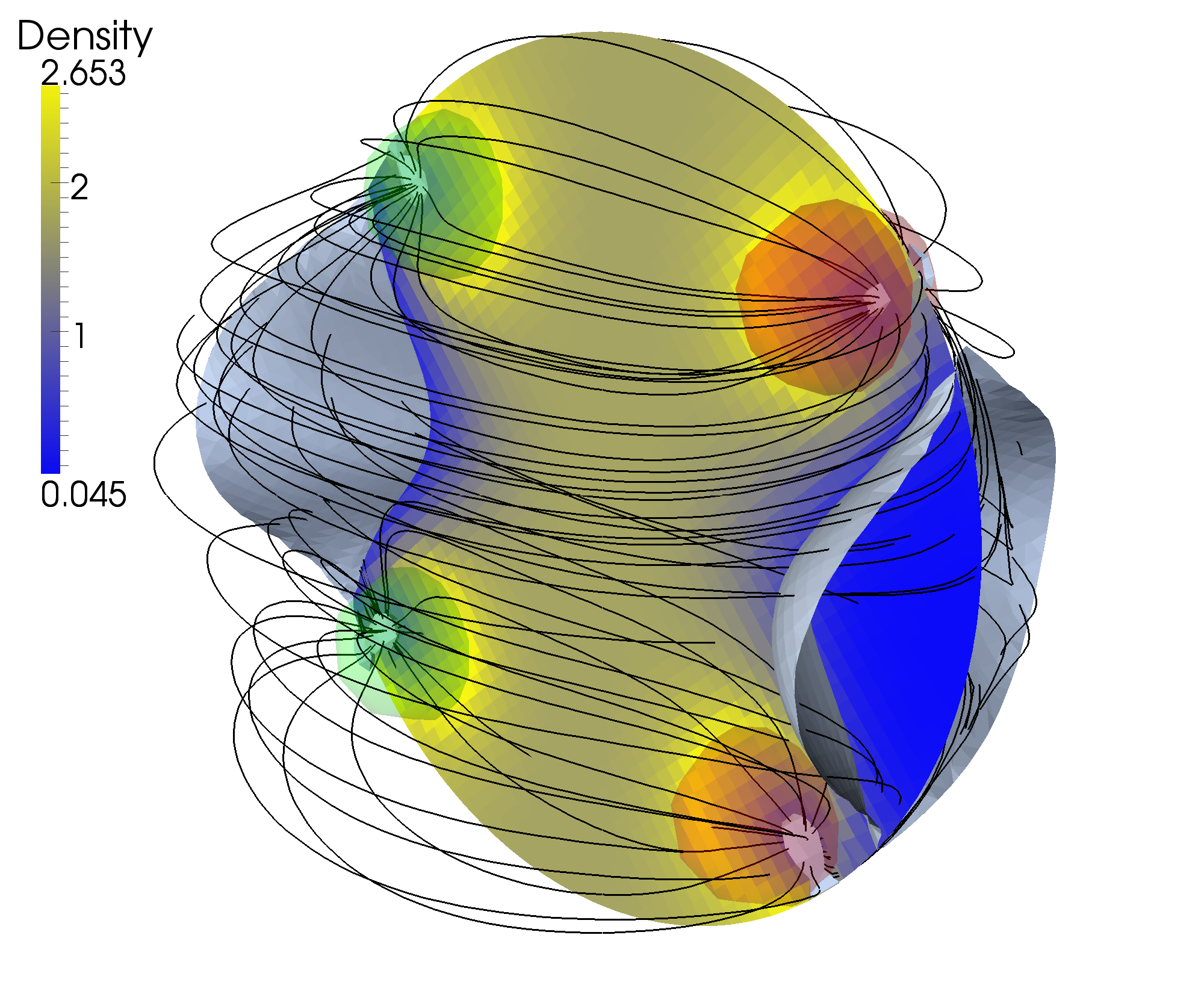}}\hspace{0cm}
		  \subfigure[~$\bar{f}=-0.08727$]{\includegraphics[width=42mm]{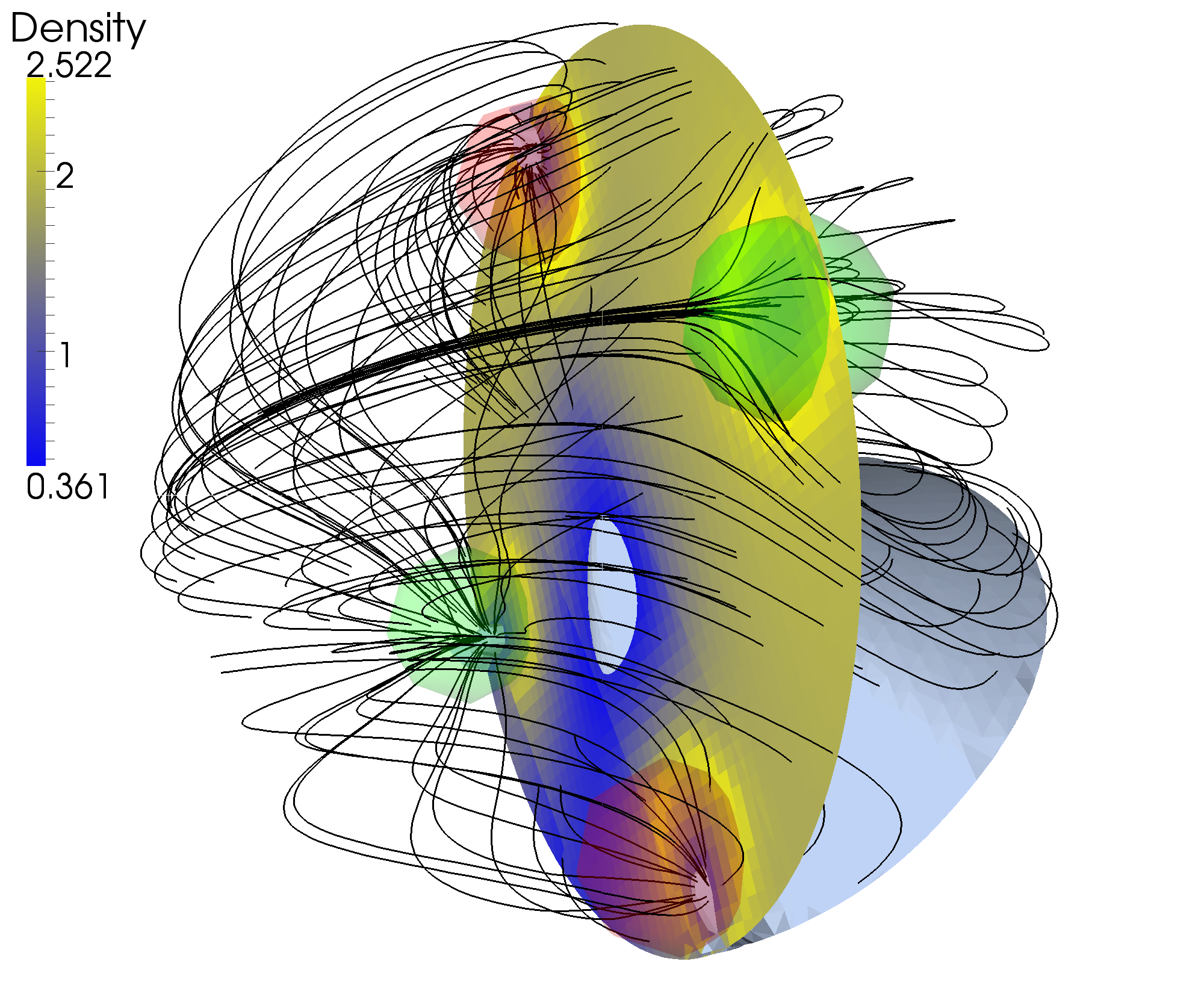}}\hspace{0cm}  }
\caption{A pair of dipolar sources, $\rho^\pm=\pm 1$, (a) parallel and (b) anti-parallel. The radius of the confining sphere is 32.
Note in (b) that the director field inside the closest green region is defect-free, yet containing a well visible splay deformation. The key is explained in Fig.\,\ref{no-sources}.}
\label{two-dipoles-sources}
\end{center}
\end{figure}

\begin{figure}
\begin{center}
	\mbox{\subfigure[]{\includegraphics[width=42mm]{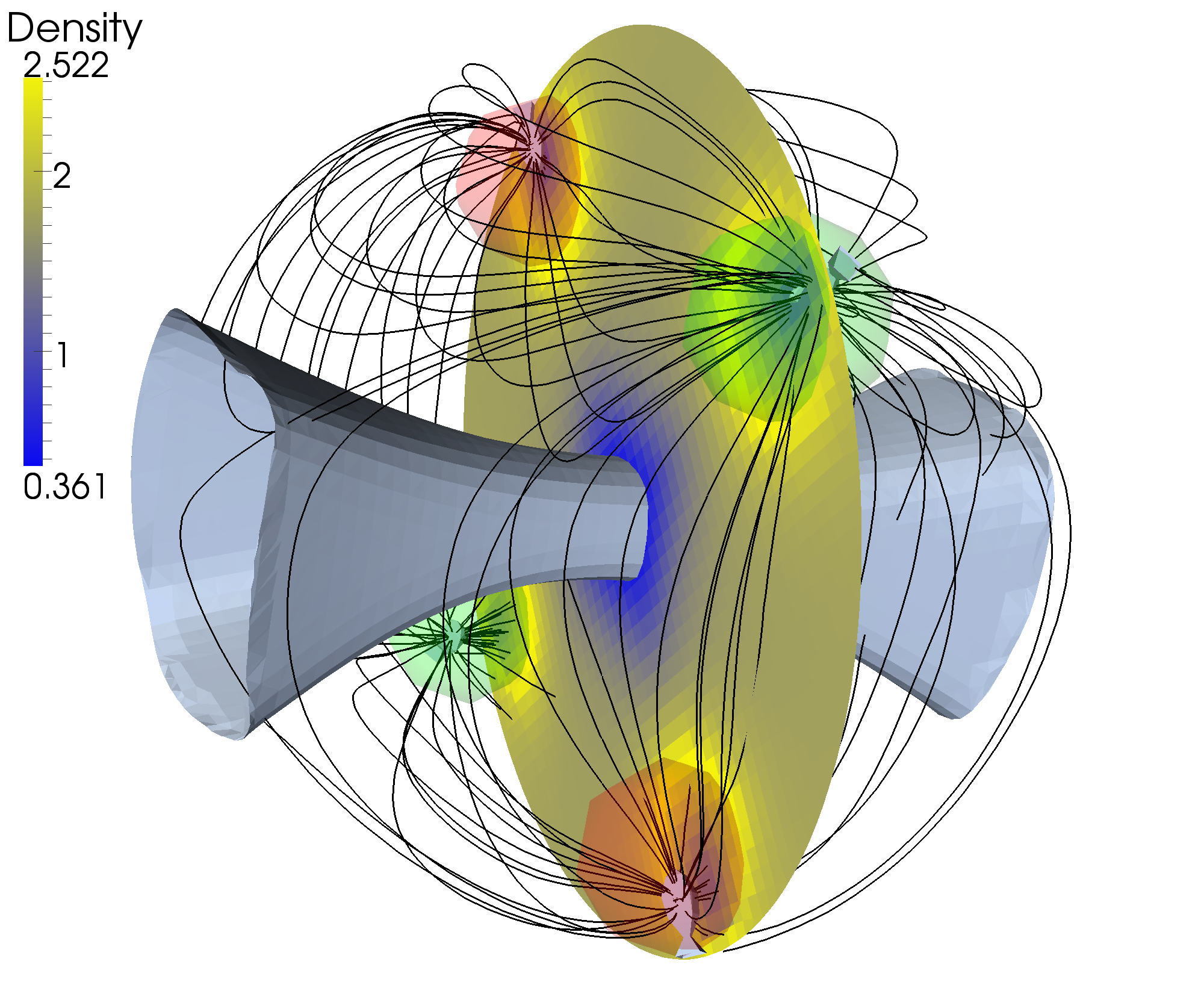}}\hspace{0cm}
		  \subfigure[~$\bar{f}=-0.08967$]{\includegraphics[width=42mm]{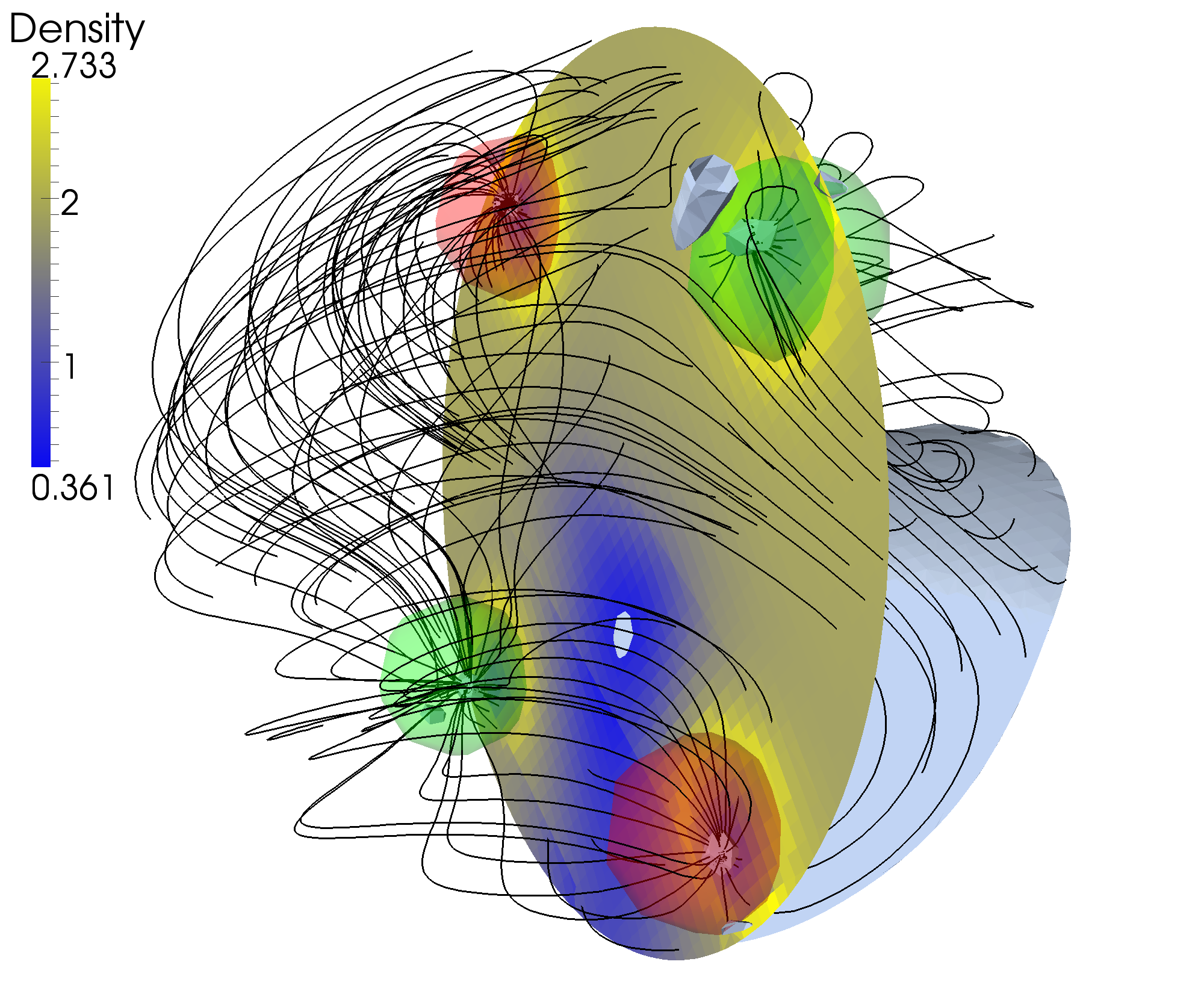}}\hspace{0cm}  }
\caption{A pair of antiparallel source dipoles, Fig.\,\ref{two-dipoles-sources}(b): (a) early stages of the evolution showing a more graspable symmetry than the actual equilibrium state in Fig.\,\ref{two-dipoles-sources}(b). In (b) the equilibrium configuration in the presence of elastic anisotropy as discussed in Sec.\,\ref{sec-anisotropy} is shown. Note that in this case all four source regions are populated by defects, in contrast to Fig.\,\ref{two-dipoles-sources}(b).
The radius of the confining sphere is 32. The key is explained in Fig.\,\ref{no-sources}.}
\label{antipar-sources}
\end{center}
\end{figure}

We have analyzed the case of no sources, the case of symmetrically and asymmetrically placed dipolar sources, and the case of a single source. We now deal with less symmetric source configurations corresponding to multiple sources.
As an example we show, Fig.\,\ref{two-dipoles-sources}, a pair of dipolar sources in parallel and antiparallel arrangements in the case of less pronounced confinement within a bigger sphere. The parallel configuration, Fig.\,\ref{two-dipoles-sources}(a), is naturally formed by two halves with two depletion regions for space filling reasons. The antiparallel configuration, Fig.\,\ref{two-dipoles-sources}(b), is less regular and seems to be quite frustrated, yet the average free energy density is only slightly larger than in the parallel case. In the early stages of the evolution, the antiparallel configuration appears to be more tractable, Fig.\,\ref{antipar-sources}(a), but is then transformed into the equilibrium state of Fig.\,\ref{two-dipoles-sources}(b).


\subsection{Effects of elastic anisotropy}
\label{sec-anisotropy}

In the calculations presented above we have assumed everywhere that $L_2=L_3=0$, i.e. we assumed a one constant approximation of the nematic elasticity. This limitation should not be too severe since splay is inherently different due to the coupling to the density, but one nevertheless needs to check whether there are any substantial variations in the equilibrium configurations if the elasticity is anisotropic, in particular in those exhibiting the depletion regions.

\begin{figure}
\begin{center}
	\mbox{\subfigure[~$\bar{f}=-0.09390$]{\includegraphics[width=42mm]{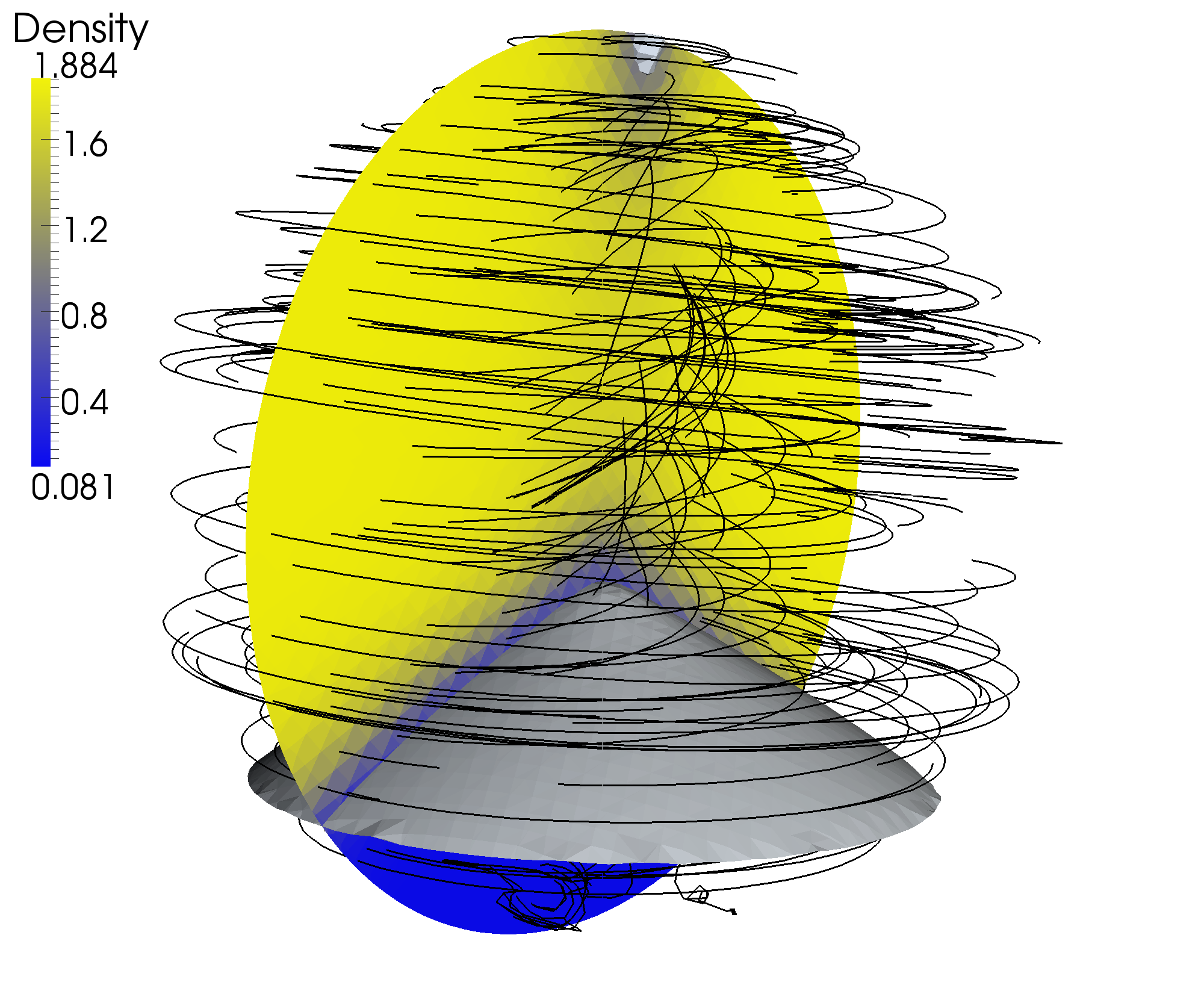}}\hspace{0cm}
		  \subfigure[~$\bar{f}=-0.09219$]{\includegraphics[width=42mm]{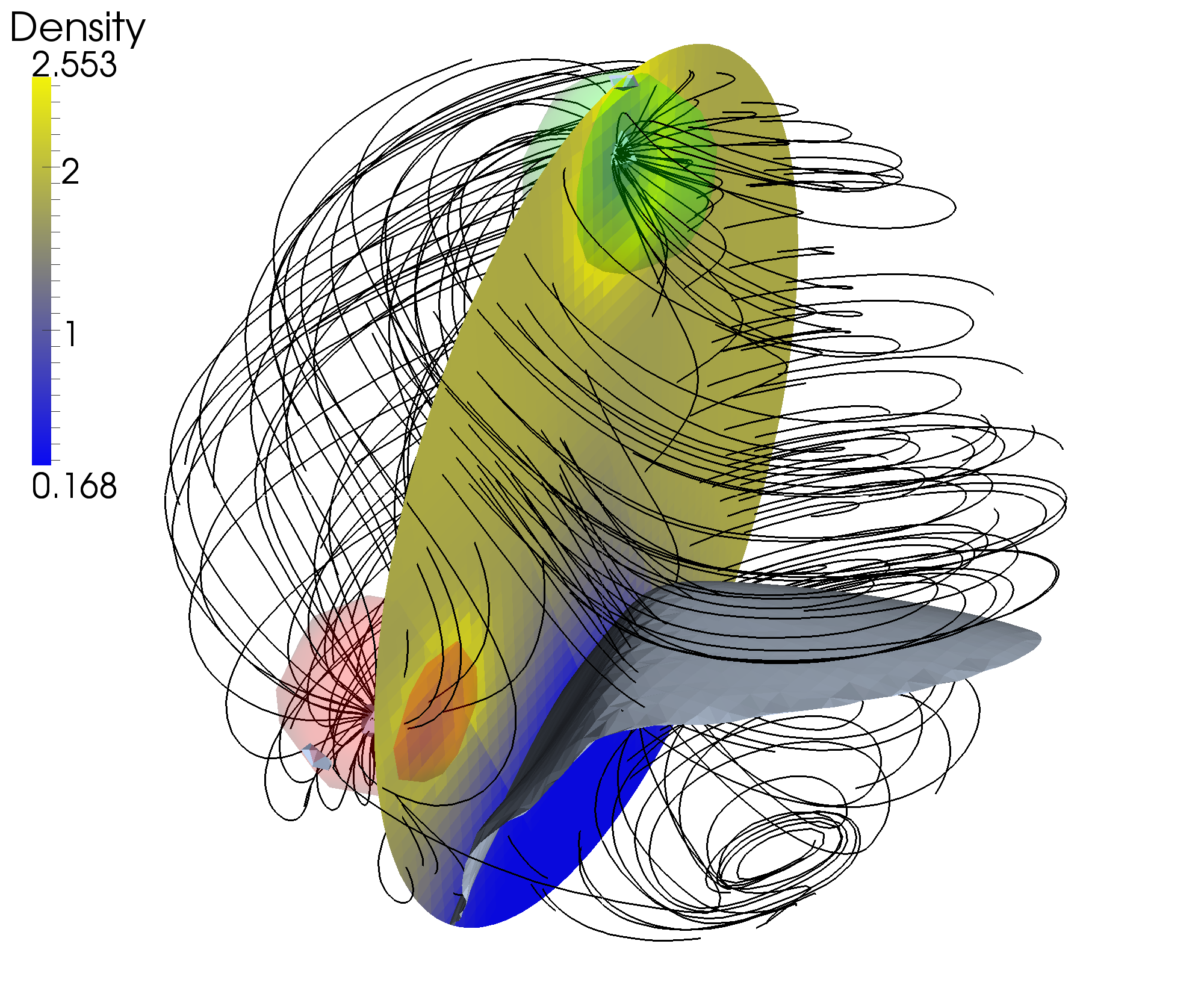}}\hspace{0cm}  }
\caption{Anisotropic Frank elasticity, $L_1=L_2=0.5$, $L_3=1$: (a) no sources, (b) oblique sources.
The radius of the confining sphere is 32. The configurations are to be compared to those on Figs.\,\ref{soft-no-sources}(b) and \ref{soft-oblique-sources}(b). The key is explained in Fig.\,\ref{no-sources}.}
\label{soft-aniso}
\end{center}
\end{figure}

We thus assume now that the elastic anisotropy is present and we fix the elastic constants to $L_1=L_2=0.5$, $L_3=1$. Obviously, Figs.\,\ref{soft-aniso} and \ref{antipar-sources}(b),  the general features of the equilibrium nematic order and density configurations presented so far, in particular the nontrivial cases with the depletion, are robust with respect to the introduction of elastic anisotropy. In Fig.\,\ref{soft-aniso} the configurations with (a) no sources and (b) oblique sources are shown  while the other parameters are the same as in Figs.\,\ref{soft-no-sources} and \ref{oblique-sources}.

Most notably, Fig.\,\ref{soft-aniso}(a), the polymer is again completely depleted from a part of the available space.
There is no strict inverse-spool ordering like in Fig.\,\ref{soft-no-sources}(b), i.e., the axial defect line is substituted for the point defect via spiraling. Such elastic-anisotropy-dependent structural transitions are well known in regular nematics \cite{zumer-crawford}.
In the case of oblique sources, Fig.\,\ref{soft-aniso}(b), the director field is much more irregular than that of Fig.\,\ref{soft-oblique-sources}(b), but the main feature -- the depletion -- is still there.
One can thus conclude that with anisotropic elasticity only details of the director field change, while the density depletions and the overall geometry of the nematic order remain the same as in the isotropic case.

\subsection{Self-consistent distribution of sources}

So far we have studied cases without or with a fixed distribution of sources. 
Let us finally check whether the distribution of sources can be determined self-consistently together with the director and density fields. In other words, $\rho^\pm({\bf r})$ will be treated as an additional variable rather than a fixed field as before. Of course, an additional penalty term for source variations must be added to the free energy density (\ref{f_bulk})-(\ref{f_gradrho}). The simplest possibility is a quadratic penalty potential 
\begin{equation}
	f_s = {1\over 2}\chi_\pm (\rho^{\pm})^2,
\end{equation}
where $\chi_\pm$ is a source density compressibility. It is required that
\begin{equation}
	\int\!\! {\rm d}V\,\rho^\pm = 0,
\end{equation}
therefore an additional Lagrange multiplier is introduced. The procedure of satisfying this constraint is fully analogous to satisfying the mass constraint, Eqs.\,(\ref{F-lambda}), (\ref{mass_conservation}), (\ref{EL_rho}), and (\ref{DeltaRho}).

\begin{figure}
\begin{center}
	\mbox{\subfigure[~$\bar{f}=-0.07727$]{\includegraphics[width=42mm]{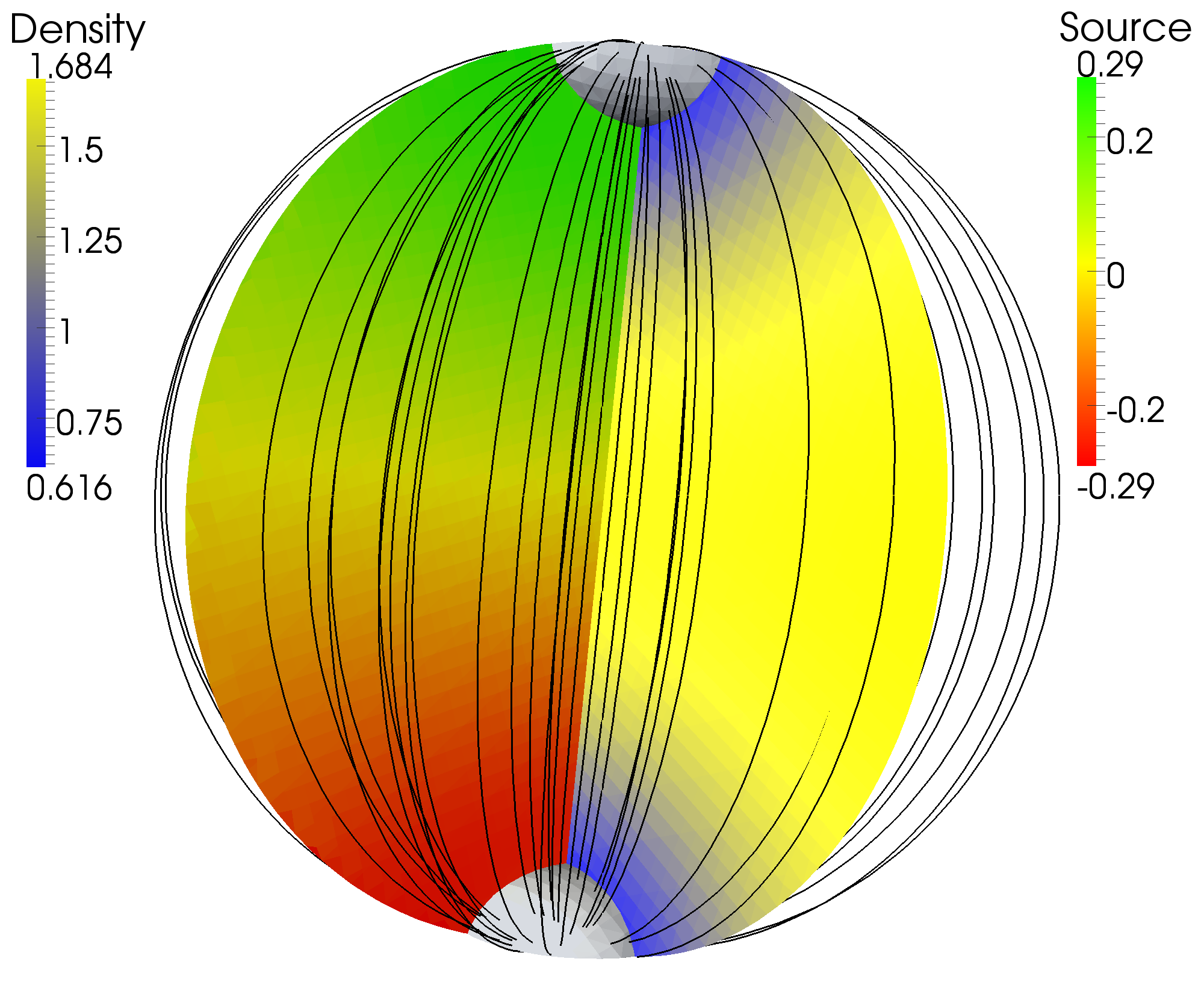}}\hspace{0cm}
		  \subfigure[~$\bar{f}=-0.09318$]{\includegraphics[width=42mm]{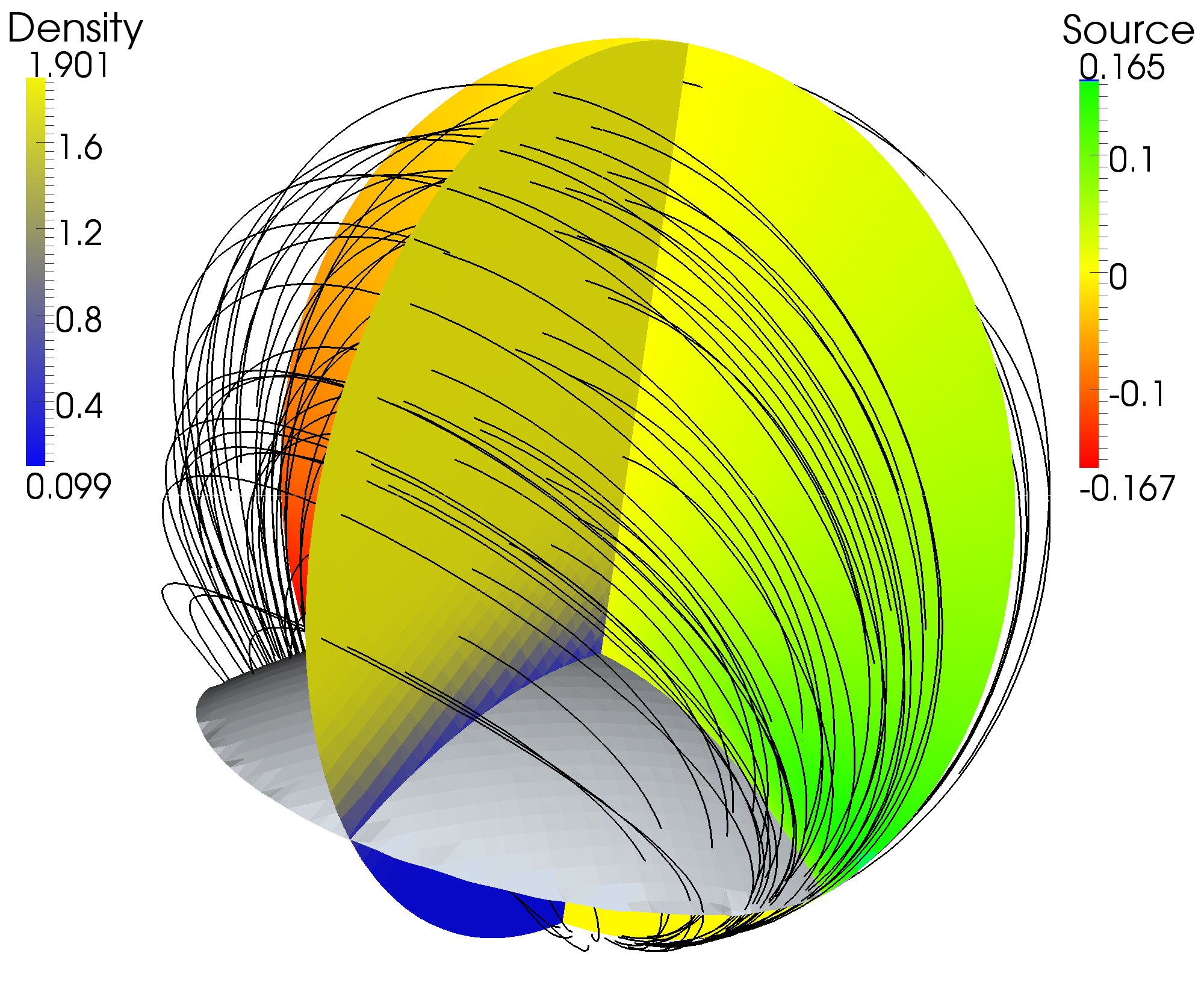}}\hspace{0cm}  }
\caption{a self-consistent source field: $\rho^\pm$ is treated as a variable; $\chi_\pm = 0.5$, density variation cheaper. 
The radius of the confining sphere is (a) 10 and (b) 32. The key is explained in Fig.\,\ref{no-sources}. The source field is shown in the additional red-green cross section.}
\label{soft-chiPM0_5}
\end{center}
\end{figure}

In the trivial case where $\chi_\pm$ is zero, the density of sources simply adjusts such that it exactly compensates for any divergence of $\rho{\bf a}$ and thus the director-density coupling energy (\ref{f_G}) is zero identically, i.e., this corresponds to putting $G=0$.
For any nonzero $\chi_\pm$, however, the source density is penalized and cannot adjust freely. The configurations obtained with $\chi_\pm = 0.5$ are shown in Fig.\,\ref{soft-chiPM0_5}. The source density is well-defined and is typically dipolar, favouring the bipolar director configuration in the small sphere, Fig.\,\ref{soft-chiPM0_5}(a), over the axially depleted one in Fig.\,\ref{soft-no-sources}(a). These two configurations can be compared directly, as all parameters, apart from the sources variation, are the same.
In the bigger sphere, Fig.\,\ref{soft-chiPM0_5}(b), one half of the bipolar director field is traded for the depleted region which absorbes the two point defects of the bipolar director configuration. The director field resembles the one in Fig.\,\ref{soft-oblique-sources}(b), but there the defects appear to be forced by the fixed localized sources to stay away from the depletion. One can verify that in Fig.\,\ref{soft-chiPM0_5}(a) and (b) the average free energy density is indeed lower than in any other configuration that can be compared directly with it.

\begin{figure}
\begin{center}
	\mbox{\subfigure[~$\bar{f}=-0.07630$]{\includegraphics[width=42mm]{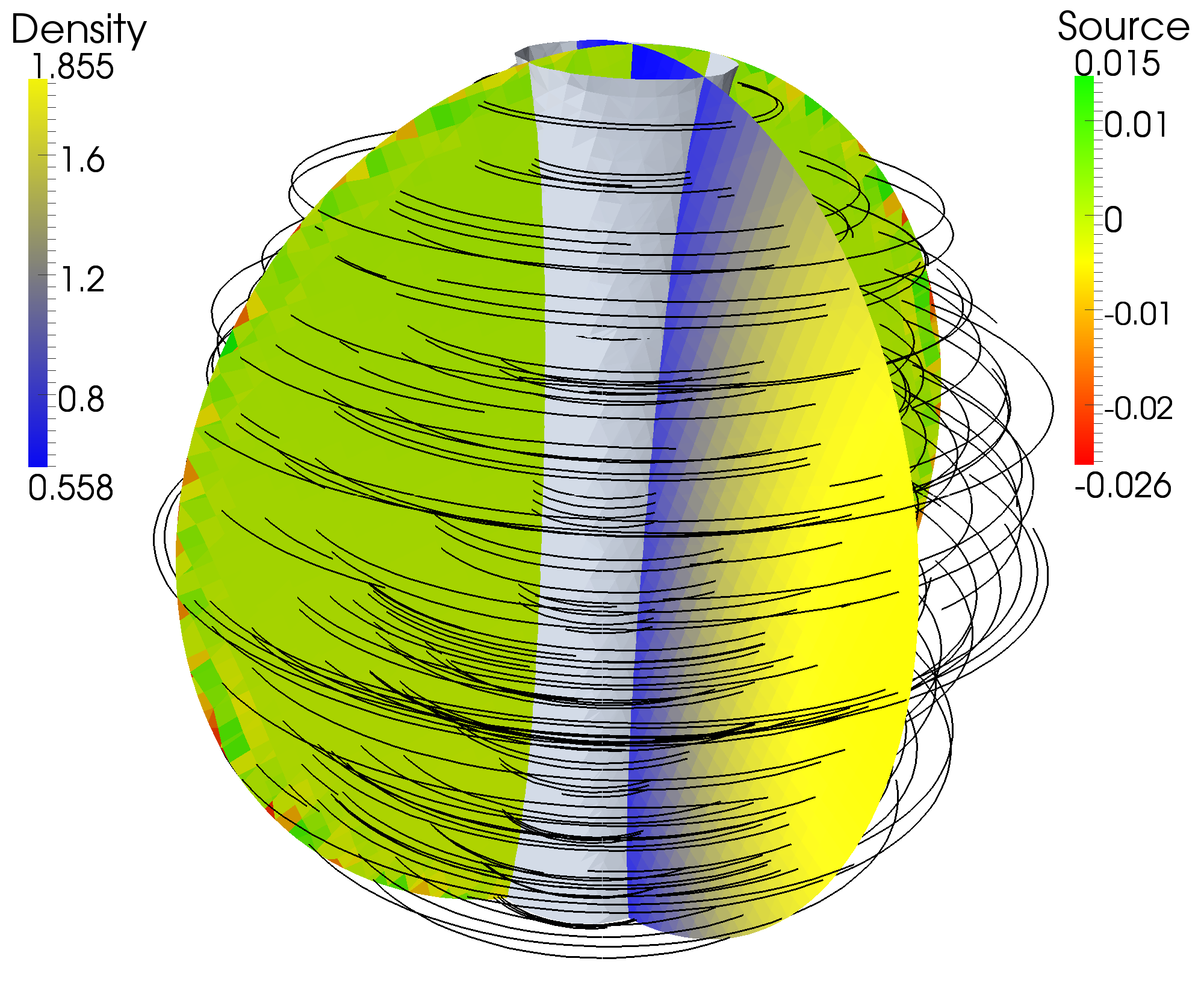}}\hspace{0cm}
		  \subfigure[~$\bar{f}=-0.09288$]{\includegraphics[width=42mm]{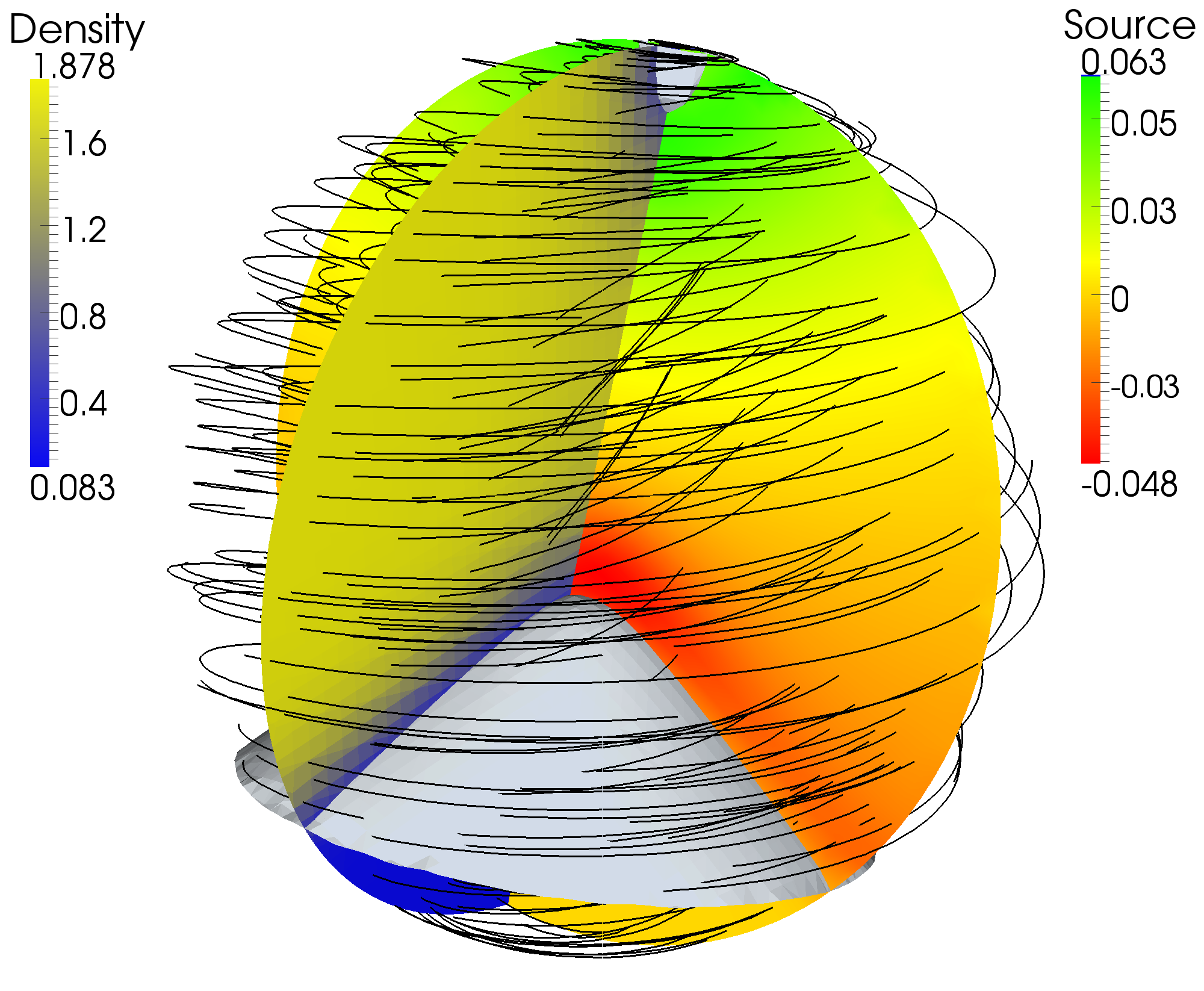}}\hspace{0cm}  }
\caption{a self-consistent source field: $\rho^\pm$ is treated as a variable; $\chi_\pm = 1$, density variation cheaper. 
The radius of the confining sphere is (a) 10 and (b) 32. The key is explained in Fig.\,\ref{no-sources}. The source field is shown in the additional red-green cross section.}
\label{soft-chiPM1}
\end{center}
\end{figure}

If, however, the source penalty gets larger, $\chi_\pm = 1$ in Fig.\,\ref{soft-chiPM1}, then the source density falls towards zero rapidly and more so for the tighter confinement. There appears to exist a structural threshold, above which the source density is zero exactly. The verification and quantification of this threshold is not within our aim at present. The configurations in Figs.\,\ref{soft-chiPM1}(a) and \ref{soft-no-sources}(a) are identical, as is the average free energy density. Moreover, the fact that the average free energy density is slightly(!) higher in Fig.\,\ref{soft-chiPM1}(a) tells us that the configuration is not yet completely relaxed to equilibrium and so the source density will fall further. On the other side, the average free energy in Fig.\,\ref{soft-chiPM1}(b) is lower than in Fig.\,\ref{soft-no-sources}(b), indicating that the source density is actually nonzero as presented in the figure. Apart from that, the configuration is very similar to Fig.\,\ref{soft-no-sources}(b).

\section{Conclusions and discussion}

Ordered structures containing a small number of director defects are typically encountered in regular nematics when subject to a tight confinement that exhibits a sufficiently strong director anchoring at the confining walls \cite{zumer-crawford}. Configurations like those in Figs.\,\ref{soft-no-sources}(a) and \ref{soft-dipolar-sources}(b) are typical for regular nematics in spherical confinement (e.g. nematic droplets). The variable density and its coupling to the splay deformation in case of nematic polymers opens the door to a variety of new features and configurations. The spontaneous appearance of depleted regions (phase separation) is particularly interesting in this context. On the other hand we must not forget that by using the polar ordering instead of the usual nematic quadrupolar ordering, we give up a class of configurations containing defects of half-integer strengths \cite{degennes}.

T5 cryoelectron microscopy experiments \cite{Amelie} show that in case of a loose packing inside
a bacteriophage capsid, the DNA toroids condensed with spermine ($sp^{4+}$) often occupy only a part of the available volume rather than being spread over the entire capsid.  Contrary to the toroids observed in the bulk \cite{Hud}, the toroids confined to a viral capsid often show no polar symmetry. The nematic and density field configurations obtained in our work, containing pronounced depletion regions,  show that the partial packing scenario could be in principle attributed already to a macroscopic free energy of the type (\ref{f_bulk})-(\ref{f_gradrho}), as an alternative to the strictly cohesive energy between DNA segments. Cohesive interaction free energies are inferred from experiments on the bulk polyvalent-counterion condensed DNA \cite{Rau-attractions,Raspaud1,Raspaud2} where the nematic order as well as the DNA density are homogeneous within the sample. In viruses the DNA-DNA cohesion mediated by polyvalent counterions such as $sp^{4+}$ \cite{Amelie,Livolant}, coupled with strong spatial confinement and nematic elasticity, might play an altogether different role leading to structures diverging from a simple inverse spool paradigm with a pronounced lack of polar symmetry. Our numerical results certainly point to the same conclusion: we obtain depleted regions without polar symmetry in the cases with larger confining spheres, Figs.\,\ref{soft-no-sources}, \ref{soft-single-source}, \ref{soft-oblique-sources}. The same conclusion regarding the absence of polar symmetry in the encapsidated toroidal aggregate holds also in the case of anisotropic elasticity, Fig.\,\ref{soft-aniso}, where the polymer is expelled from one polar region but not from the other. We note here that our simplified macroscopic free energy, (\ref{f_bulk})-(\ref{f_gradrho}), can not capture all the subtleties of the intersegment interactions of a real nematic polymer such as DNA \cite{osm1,osm2}.

Recent elucidation of the organization of the DNA packing in the T5 capsid at various DNA densities as it undergoes a series of phase transitions on decreasing the DNA density \cite{domains} attests to the possible minor role of inverse spool states in viral DNA packing \cite{Francoise}. Though our results definitely point to the existence of states with partial depletion regions in polymer packing with no polar symmetry, they also indicate that the inverse spool family of states is nevertheless quite robust within our model assumptions. This finding is completely consistent also with recent molecular simulations, see below \cite{Angelescu}. Strong confinement for smaller confining spheres appears to induce toroidal packing irrespective of all the other terms in the free energy
and  points to a {\sl universality} of the inverse spool configuration. The observed non-inverse spool states might thus originate in other types of interaction energy e.g. between the DNA and various molecular moieties or charged groups distributed along the inner surface of the capsid, effects which are not considered in this work. 

Experiments on DNA ejection from partially filled capsids also shows a profusion of monodomains of nematic order separated by dislocation, twist and bend walls \cite{domains}. Though we have explored in relative detail the parameter space of our numerical solutions of confined polymer nematic packing we were unable to detect packings with domain structure of the type observed in these experiments. The same is true for the experimentally observed sequence of liquid crystalline phases on ejection of DNA, starting with hexagonal and ending in isotropic \cite{domains}. Our free energy and nematic order description is of course insufficient for the description of these effects. A serious limitation of our model might be the simplified form of the density dependence of the free energy which is basically reduced to the quadratic deviation term, Eq.\,(\ref{f_rho}). The osmotic pressure measured directly for bulk DNA \cite{osm1,osm2} certainly shows the inadequacy of this simplification and would in fact lead to a very complicated density variation of the free energy with pronounced salt and polyvalent counterion concentration dependence. Unfortunately the details of this dependence are not known for a wide range of parameters and are understood even less on a molecular level. Our choice for a simplified version of the density dependence is thus in this respect fully vindicated by the blind spots in our understanding of the fundamental interactions between DNA molecules. It seems appropriate that theoretical formulations of the DNA viral packing problem remain as close to the experimentally determined variables as possible \cite{siber}.

Our numerical solution of the confined polymer nematic model also provides an interesting alternative to simulations of DNA packing within icosahedral viruses \cite{Angelescu}. Since the seminal simulation work of Kindt et al. \cite{kindt}  inverse spool-like structures have been obtained on various levels of sophistication. Typically in a simulation a semiflexible chain, possibly with chiral twist interactions as in the work of  Spakowitz and Wang \cite{spakowitz}, is endowed with different types of self-interaction that contains repulsive screened electrostatic and/or attractive ion-correlation or hydration contributions. The DNA packaging starts at the inner capsid surface, typically with a helical winding along the inner surface of the sphere-like capsid \cite{Slosar,Angelescu2}, and then proceeds inward towards a more disordered core using all available inner space \cite{arsuaga,LaMarque,spakowitz}. Sometimes the capsid is modeled with more detail than in the spherical model, including also icosahedral symmetry of the viral capsids \cite{Angelescu3}. Details of the interaction potential also seem to be very important. Presence of intersegment attractions in DNA-DNA interactions has been shown to lead to configurations that do not conform to the inverse-spool paradigm. Forrey and Muthukumar were the first to observe a folded toroid state in DNA packing with attractive self-interactions within an icosahedral enclosure \cite{Forrey}. The same packing configuration was later seen also in \cite{Petrov}. The spools and folded toroids observed in these simulations are an outcome of a subtle interplay between chain flexibility, size of the enclosure and steric interactions between DNA segments. It appears that the increase of the persistence length which would amount to the same thing as decreasing the size of the enclosure would promote the transition from folded toroid to a spool-like structure \cite{Forrey}.

It thus seems that several salient features of the simulation results are in accord with our mesoscopic modeling of the confined polymer nematic ordering. First of all the robustness of the inverse spool configuration transpires from both types of approaches and points to its possible universality. Though ordered states discussed in this work never seem to point to the existence of folded spool configurations we do get other types of related ordering with pronounced depleted regions and no polar symmetry. 

Though one gets the impression that molecular simulations of nematic polymer packing can incorporate more relevant details of molecular interactions and chain flexibility, it should be stated quite clearly that the details of DNA-DNA interactions in particular are still not quantitatively known with great precision \cite{osm1}. The mechanism of polyvalent-counterion mediated attraction keeps defying proper understanding since apart from the charge of the counterions \cite{Kanduc} it appears to be related also to ion-specific non-electrostatic interactions that are notoriously difficult to parameterize and model \cite{Yaakov}. The strong point of mesoscopic modeling as pursued in this work is that at least in principle one can include the experimentally measured density dependence of the free energy into the calculation directly by generalizing the quadratic term, Eq.\,(\ref{f_rho}), without invoking any particular microscopic mechanism leading to the experimentally observed equation of state. This type of generalizations will be pursued in the future.

\begin{acknowledgements}
This work has been supported by the Agency for Research and Development of Slovenia under Grants No. P1-0055 and
No. J1-0908.
\end{acknowledgements}

\end{document}